\documentclass[journal]{IEEEtran}
\usepackage{cite}

\ifCLASSINFOpdf
  \usepackage[pdftex]{graphicx}
\else
  \usepackage[dvips]{graphicx}
\fi
\usepackage[cmex10]{amsmath}
\usepackage{amssymb}
\usepackage{algorithmic,algorithm}

\usepackage{array}

\ifCLASSOPTIONcompsoc
 \usepackage[caption=false,font=normalsize,labelfont=sf,textfont=sf]{subfig}
\else
 \usepackage[caption=false,font=footnotesize]{subfig}
\fi
\usepackage{mathtools}
\usepackage{bm}
\usepackage{booktabs}
\usepackage{tabto}
\usepackage{hyperref}

\newtheorem{theorem}{Theorem}
\newtheorem{proposition}{Proposition}
\newtheorem{corollary}{Corollary}

\DeclareMathOperator*{\argmin}{arg\,min}
\DeclareMathOperator*{\argmax}{arg\,max}

\newcommand{\Lp}[1]{\mathcal{L}_{\mu}^#1}
\newcommand{\remidx}[2]{\bm{#1}_{[#2]}}
\newcommand{\pmin}{p'}
\newcommand{\pmax}{p''}

\newcommand{\Pband}{\mathcal{P}^{=}}
\newcommand{\dint}{\, \mathrm{d}}
\newcommand{\prox}[2]{\textbf{prox}_{#1}(#2)}

\makeatletter
\newcommand{\subalign}[1]{%
  \vcenter{%
    \Let@ \restore@math@cr \default@tag
    \baselineskip\fontdimen10 \scriptfont\tw@
    \advance\baselineskip\fontdimen12 \scriptfont\tw@
    \lineskip\thr@@\fontdimen8 \scriptfont\thr@@
    \lineskiplimit\lineskip
    \ialign{\hfil$\m@th\scriptstyle##$&$\m@th\scriptstyle{}##$\crcr
      #1\crcr
    }%
  }
}
\makeatother

\begin{document}

\title{On the Minimization of Convex Functionals of Probability Distributions Under Band
       Constraints}

\author{Michael~Fau\ss{},~\IEEEmembership{Member,~IEEE}
        and~Abdelhak~M.~Zoubir,~\IEEEmembership{Fellow,~IEEE}
\thanks{M.~Fau\ss{} and A.~M.~Zoubir are with the Signal Processing Group, Institute for
        Telecommunications, Department of Electrical Engineering and Information Technology,
        Technische Universit\"at Darmstadt, Merckstr. 25, 64283 Darmstadt, Germany. E-mail:
        \{fauss,zoubir\}@spg.tu-darmstadt.de.}%
\thanks{Manuscript received March 20, 2017.}}

%
%

\markboth{IEEE Transactions on Signal Processing, Vol.~XX, No.~XX, Month XX, 20XX}%
{Fau\ss{} and Zoubir: On the Minimization of Convex Functionals of Probability Distributions Under Band Constraints}

%



\maketitle

\begin{abstract}
  The problem of minimizing convex functionals of probability distributions is solved under the assumption that the density of every distribution is bounded from above and below. A system of sufficient and necessary first-order optimality conditions as well as a bound on the optimality gap of feasible candidate solutions are derived. Based on these results, two numerical algorithms are proposed that iteratively solve the system of optimality conditions on a grid of discrete points. Both algorithms use a block coordinate descent strategy and terminate once the optimality gap falls below the desired tolerance. While the first algorithm is conceptually simpler and more efficient, it is not guaranteed to converge for objective functions that are not strictly convex. This shortcoming is overcome in the second algorithm, which uses an additional outer proximal iteration, and, which is proven to converge under mild assumptions. Two examples are given to demonstrate the theoretical usefulness of the optimality conditions as well as the high efficiency and accuracy of the proposed numerical algorithms.
\end{abstract}

\begin{IEEEkeywords}
  Robust statistics, distributional uncertainties, band model, convex optimization, block coordinate descent, $f$-divergence 
\end{IEEEkeywords}

%
\IEEEpeerreviewmaketitle

\section{Introduction}
\label{sec:intro}
%
%
%
%
\IEEEPARstart{F}{unctionals} of probability distributions play a central role in probability theory and statistics. To clarify, a functional is a mapping from a vector space to the real line, i.e., a function which maps an element of a possibly high- or infinite-dimensional space to a scalar value. Omnipresent examples are the moments of a real-valued random variable, which map a distribution to a real number.

Convex functionals of probability distributions occur naturally in problems of statistical inference and decision making. In general, the expected cost of any inference procedure with a convex loss function can be shown to be a convex functional of probability distributions \cite{Nguyen2009}. Consequently, examples of convex functionals can be found in detection \cite{Levy2008}, estimation \cite{Klebanov1978}, and joint detection and estimation theory, as well as in Bayesian inference \cite{Reid2011}.

In practice, it is often the case that the distributions of random variables that describe a random phenomenon are not known exactly, but are subject to uncertainty. This uncertainty can, for example, be caused by a lack of information about the random phenomenon or the absence of an appropriate model for its mathematical description. Uncertainty can also be introduced intentionally in order to safeguard an inference procedure against deviations from the underlying assumptions. The type and degree of uncertainty is usually specified by means of an \emph{uncertainty set} that contains all feasible distributions. Given such a set, a question that naturally arises is which distributions are \emph{most favorable} and which are \emph{least favorable}. In order to answer this question, a given cost function has to be minimized or maximized over the uncertainty set. This is the problem addressed in this paper.

The uncertainty model that is assumed to hold throughout the paper restricts the densities of feasible distributions to lie within a band that is defined by two non-intersecting functions, which bound the density from above an below. It can be thought of as a confidence interval for the true density function. This model is known as the \emph{density band model} and is commonly used in robust statistics \cite{KassamPoor1985}. In the context of robust hypothesis testing it was first studied in \cite{Kassam1981} and was recently revisited in \cite{Fauss2015}. It is discussed in more detail in Section~\ref{sec:band_model}. 

The problem that motivated the work in this paper relates to the design of minimax optimal sequential tests for multiple hypotheses. The least favorable distributions for this type of test can be shown to be minimizers of functionals of the form\footnote{
  For binary tests, this result can be found in \cite{Fauss2016_thesis}. The corresponding results on tests for multiple hypotheses are to be presented in a forthcoming publication.
}
\begin{equation}
  \int_{\Omega} f\biggl(\omega,\frac{dP_1}{dP}(\omega),\ldots,\frac{dP_N}{dP}(\omega)\biggr) \dint P(\omega),
  \label{eq:motivation_functional}
\end{equation}
where $P,P_1,\ldots,P_N$ denote distributions that are subject to uncertainty and the function $f$ is jointly convex in the likelihood ratios $d P_n/d P$. A particular difficulty that arises in the design of minimax sequential tests is that $f$ is itself the solution of an optimization problem so that its value and its derivatives can only be evaluated numerically. Minimizing \eqref{eq:motivation_functional} over $P,P_1,\ldots,P_N$, and under band constraints, is a challenging task because it involves two nested optimization problems. Deriving analytic solutions, or determining approximate solutions is usually not possible. Therefore, a suitable numerical algorithm is required that is:
\begin{itemize}
  \item accurate enough to closely approximate continuous density functions,
  \item efficient enough to handle multiple distributions and reasonably fine grids for their discretization;
  \item parallelizable, in order to leverage modern hardware;
  \item and robust against mild numerical noise in the evaluation of $f$ and its derivatives.
\end{itemize}
Looking into existing convex optimization frameworks and algorithms, it was found that most off-the-shelf methods did not satisfy these requirements. On the one hand, high performance solvers such as Gurobi, MOSEK, or CPLEX were found to be too restrictive in terms of feasible objective functions. On the other hand, commonly used generic convex optimization algorithms, such as interior point, steepest descent or conjugate gradient methods, turned out to be too inefficient to be useful in practice. In addition, all algorithms considered in the survey suffered from severe accuracy issues, especially in the tails of the optimal distributions.

In this work, an approach for the minimization of convex functionals of probability distributions, under density band uncertainty, is detailed, which is efficient, reliable, offers control over the achieved accuracy, and is applicable beyond the particular use case of robust sequential hypothesis testing. The proposed algorithms offer a good trade-off between being generic and specific, in the sense that they heavily exploit assumptions about the structure of the objective function and the constraints, while at the same time providing enough flexibility to be applicable to a large class of problems in statistical signal processing and robust statistics in particular.

The existing literature on convex functionals is large and dispersed. While in the functional analysis literature the term convex functionals prevails \cite{Rockafellar1968}, in statistics, signal processing, and information theory, similar classes of functions often go by the names divergence, distance, dissimilarity, or disparity. Early results on the subject are due to, among others, Pearson \cite{Pearson1900}, Mahalanobis \cite{Mahalanobis1930}, Shannon \cite{Shannon1948},
and Kullback \cite{Kullback1951}---see \cite{Liese1987,Pardo2005} and references therein for a detailed treatment. The minimization of convex functionals, and their relation to robust decision making, has been addressed by Huber \cite{Huber1973}, Poor \cite{Poor1980}, Kassam \cite{Kassam1981}, and Guntuboyina \cite{Guntuboyina2011} to name just a few. An approach similar to the one presented here was used in \cite{DAmico2014} to solve a constrained minimization problem with a separable cost function. However, neither is the problem in \cite{DAmico2014} a special case of the problem investigated in this paper, nor vice versa. 

The paper is organized as follows: the density band uncertainty model is briefly reviewed in Section~\ref{sec:band_model}. In Section~\ref{sec:problem_formulation}, the functional minimizaton problem is stated in a formal manner. Necessary and sufficient optimality conditions as well a bound on the optimality gap of a feasible candidate solution are stated in Section~\ref{sec:optimality_conditions_and_gap}. Both are useful results in their own right and constitute a main contribution of the paper. In Section~\ref{sec:calculation}, the functional optimization problem is discretized and an algorithm is proposed that iteratively solves the optimality conditions using a block coordinate descent (BCD) strategy. Since the latter is not guaranteed to converge to a global minimum for general convex objective functions, a second algorithm is introduced that augments the objective function with an additional proximal term. Guaranteed convergence is shown for the proximal algorithm. In Section~\ref{sec:examples}, two examples are provided to illustrate how the optimality conditions as well as the proposed numerical algorithms can be used in practice. Section~\ref{sec:conclusion} concludes the paper.      

\emph{Notation}: Probability distributions are denoted by upper case letters, their densities by the corresponding lower case letters. Boldface lower case letters $\bm{x}$ are used to indicate row vectors and boldface upper case letters $\bm{X}$ to indicate matrices. The notations $\remidx{x}{n}$ and $\remidx{X}{n}$ are used to denote a vector whose $n$th element has been removed and matrices whose $n$th row has been removed, respectively. The inner product of two vectors $\bm{x}$ and $\bm{y}$ is denoted by $\langle \bm{x}, \bm{y} \rangle$, the element-wise product by $\bm{xy}$. All comparisons between vectors are defined element-wise. The all-ones vector is denoted by $\bm{1}$. In the pseudocode of the algorithms, an R-style arrow notation $x \leftarrow y$ is used to assign a value $y$ to a variable $x$. $\mathcal{L}_{\mu}^p$ denotes the space of functions whose $p$th power of the absolute value is integrable with respect to the measure $\mu$. The operators $(f)^+$ and $(f)^-$ denote the positive and negative parts of a function $f$, i.e., $(f(\omega))^+ = \max\{ f(\omega), 0\}$ and $(f(\omega))^- = \min\{ f(\omega), 0\}$. The notation $f(\bm{\omega})$ is used as shorthand for the vector $(f(\omega_1), \ldots, f(\omega_N))$. For functions that are defined directly on the sample space, the explicit argument $\omega$ is often omitted for the sake of a more compact notation, in particular if the function is integrated or is itself an argument of a higher-order function. Comparisons between functions are defined point-wise. The notation $\partial_{x_n} f(\bm{x})$ is used for the partial subdifferential of a convex function $f \colon \mathcal{X} \subset \mathbb{R}^N \to \mathbb{R}$ with respect to $x_n$ at $\bm{x}$, i.e.,
\begin{equation*}
  \partial_{x_n} f(\bm{x}) \coloneqq \left\{ \gamma \in \mathbb{R} : 
  \frac{f(\bm{y})-f(\bm{x})}{y_n-x_n} \leq \gamma \quad \forall \bm{y} \in \mathcal{X} \setminus \{\bm{x}\} \right\}.
\end{equation*}
Finally, the generalized inverse \cite{Embrechts2013} of a nondecreasing function $f \colon \mathcal{X} \subset \mathbb{R} \to \mathbb{R}$ is defined as
\begin{equation}
  f^{-1}(c) = \inf \left\{\, x \in \mathcal{X} : f(x) \geq c \,\right\}.
  \label{eq:f_inverse}
\end{equation}
Note that this implies that $f^{-1}$ is nondecreasing in $c$ and that for $\mathcal{X} = \mathbb{R}$
\begin{align*}
  f^{-1}(c) &= \infty,  & &\text{if} & f(x) &< c \quad \forall x \in \mathbb{R}, \\
  f^{-1}(c) &= -\infty, & &\text{if} & f(x) &> c \quad \forall x \in \mathbb{R}.
\end{align*}

\section{The Density Band Uncertainty Model}
\label{sec:band_model}

Let $(\Omega,\mathcal{F})$ be a measurable space and let $\mu$ be an absolutely continuous $\sigma$-finite measure on this space. The density-band uncertainty model specifies sets of the form
\begin{equation}
  \Pband = \{\, p \in \Lp{1}: \pmin \leq p \leq \pmax \,\},
  \label{eq:band_model}
\end{equation}
where
\begin{equation}
  0 \leq \pmin \leq \pmax \leq \infty, \quad \int_\Omega \pmin \dint \mu  \leq 1, \quad \int_\Omega \pmax \dint \mu \geq 1.
  \label{eq:band_conditions}
\end{equation}
Thus, all feasible densities are upper bounded by $\pmax$ and lower bounded by $\pmin$. The case where the upper bound is infinity and the lower bound is zero is the unconstrained case.

The density band uncertainty model is useful for several reasons. First, it provides a great amount of flexibility to the designer of an inference procedure, as it allows for varying local degrees of uncertainty on different regions of the sample space. Depending on the application, it can be constructed by hand, based on expert knowledge, or statistically, via confidence interval estimators. Second, in contrast to many parametric uncertainty models, it provides clear visualization and easy interpretation. Third, from a theoretical point of view, the density band model is of interest because it generalizes several popular uncertainty models such as the $\varepsilon$-contamination model \cite{Huber1965} and the bounded distribution function model \cite{Oesterreicher1978}. A more detailed discussion of the band model and its properties can be found in \cite{Kassam1981} and \cite{Fauss2015}.

\section{Problem Formulation}
\label{sec:problem_formulation}

Let $N$ be a positive integer. The functionals considered in this work are of the form
\begin{equation}
  I_f(p_1,\ldots,p_N) \coloneqq \int_{\Omega} f(\omega,p_1(\omega),\ldots,p_N(\omega)) \dint \mu(\omega),
  \label{def:functional}
\end{equation}
where $p_1, \ldots, p_N$ are probability densities on $(\Omega,\mathcal{F})$ and
\begin{equation}
  \begin{aligned}
    f\colon \Omega \times [0,\infty)^N &\to (-\infty,\infty]  \\
    (\omega,x_1,\ldots,x_N) &\mapsto f(\omega,x_1,\ldots,x_N)
  \end{aligned}
  \label{def:f}
\end{equation}
is a function that is convex with respect to $(x_1,\ldots,x_N)$. In order to guarantee that \eqref{def:functional} is well defined, it is further assumed that $f(\omega,p_1(\omega),\ldots,p_N(\omega))$ is $\mu$-measurable for all feasible densities $p_1,\ldots,p_N$. A proof, and a more in-depth analysis, of the existence and well-definedness of \eqref{def:functional} is detailed in \cite{Rockafellar1968}. Also note that $f$ being convex with respect to $(x_1,\ldots,x_N)$ does not imply that $f(\omega, p_1(\omega), \ldots, p_N(\omega))$ is convex with respect to $\omega$. To facilitate compact notation, the arguments of $f$ are occasionally written in vector notation and the direct dependence on $\omega$ is omitted, i.e.,
\begin{equation*}
  f(\bm{x}) \coloneqq f(x_1,\ldots,x_N) \coloneqq f(\omega,x_1,\ldots,x_N).
\end{equation*}
It is important to note that $I_f(p_1,\ldots,p_N)$ is used instead of $I_f(P_1,\ldots,P_N)$. The latter notation is commonly used in the context of distance measures between distributions, such as $f$-divergences \cite{AliSilvey1966} and $f$-dissimilarities \cite{GyorfiNemetz1977}, in order to emphasize that the distance does not depend on how the reference measure $\mu$ is chosen. However, since $f$ in \eqref{def:f} is not assumed to be homogeneous and is allowed to directly depend on $\omega$, this independence does not hold in general.

The optimization problem considered in this paper is
\begin{equation}
  \min_{\{p_n \in \Pband_n\}_{n = 1}^N} \; I_f(p_1,\ldots,p_N),
  \label{prob:min_implicit}
\end{equation}
where all $\Pband_n$ are of the form \eqref{eq:band_model} and $\{p_n \in \Pband_n\}_{n = 1}^N$ is used as a shorthand notation for $p_n \in \Pband_n$ for all $n = 1, \ldots, N$. Expressing the objective function and constraints explicitly, \eqref{prob:min_implicit} becomes
\begin{gather}
  \min_{ \{ p_n \in \Lp{1} \}_{n=1}^N } \; \int_{\Omega} f(\omega, p_1(\omega), \ldots, p_N(\omega) ) \dint\mu(\omega)
  \label{prob:min_explicit} \\
  \text{s.t.} \quad \pmin_n \leq p_n \leq \pmax_n, \quad \int_{\Omega} p_n \dint\mu = 1, \quad n = 1, \ldots, N. \notag
\end{gather}
In the next section, a sufficient and necessary condition for a density vector $\bm{q} = (q_1,\ldots,q_N)$ to be a solution of \eqref{prob:min_explicit} is given as well as a bound on the optimality gap $I_f(\bm{p})-I_f(\bm{q})$ for a feasible density vector $\bm{p}$.

\section{Optimality Conditions and Optimality Gap}
\label{sec:optimality_conditions_and_gap}

The optimality conditions for \eqref{prob:min_explicit} are given in the following theorem.

\begin{theorem}[Optimality Conditions]
  A sufficient and necessary condition for the densities $\bm{q} = (q_1,\ldots,q_N)$ to be a solution of \eqref{prob:min_explicit} is that they satisfy
  \begin{equation*}
    q_n = \begin{dcases}
            \pmax_n,                      & f_n(\bm{q}) < c_n \\
            f_n^{-1}(\remidx{q}{n},c_n),  & f_n(\bm{q}) = c_n \\
            \pmin_n,                      & f_n(\bm{q}) > c_n
          \end{dcases}
  \end{equation*}
  for some $c_1,\ldots,c_N \in \mathbb{R}$ and all $n = 1,\ldots,N$. Here $f_n$ denotes a partial subderivative of $f$ with respect to $x_n$, i.e.,
  \begin{equation*}
    f_n(\bm{x}) \in \partial_{x_n}f(\bm{x}),
  \end{equation*}
  and $f_n^{-1}(\remidx{x}{n},c)$ denotes the generalized inverse of $f_n(\bm{x})$ with respect to $x_n$ in the sense of \eqref{eq:f_inverse}.
  \label{th:optimality_condition}
\end{theorem}

A proof for Theorem~\ref{th:optimality_condition} is detailed in Appendix~\ref{apx:proof_of_optimality_condition}.

Making use of the fact that $f_n$ is a subderivative of a convex function and, hence, is nondecreasing, Theorem~\ref{th:optimality_condition} can further be written in an alternative, more expressive form that eliminates the need for the explicit case-by-case definition.

\begin{corollary}[Optimality Conditions]
  A sufficient and necessary condition for the densities $\bm{q} = (q_1,\ldots,q_N)$ to be a solution of \eqref{prob:min_explicit} is that they satisfy
  \begin{equation}
    q_n = \min \{\, \pmax_n \,,\, \max \{\, f_n^{-1}(\remidx{q}{n},c_n) \,,\, \pmin_n \,\} \,\}
    \label{eq:optimality_condition}
  \end{equation}
  for some $c_1,\ldots,c_N \in \mathbb{R}$ and all $n = 1,\ldots,N$.
  \label{cl:optimality_condition}
\end{corollary}

Corollary~\ref{cl:optimality_condition} is proved in Appendix \ref{apx:proof_of_corollary}. It gives an expression for $q_n$ solely in terms of the remaining optimal densities $\remidx{q}{n}$ and the scalar $c_n$. That is, knowing $\remidx{q}{n}$, the missing density $q_n$ can be found via a search over $c_n$. The iterative algorithms presented in the next section are based on this idea. It can further be seen from \eqref{eq:optimality_condition} that $q_n$ is a projection of $f_n^{-1}(\remidx{q}{n},c_n)$ onto the band of feasible densities $\Pband_n$. In the limit, i.e., $\pmax_n \to \infty$ and $\pmin_n \to 0$, it follows that $q_n = f_n^{-1}(\remidx{q}{n},c_n)$. The example in Section~\ref{ssec:weighted_kl_div} demonstrates this.

In practice, it might not be possible to find an exact analytic solution to the optimality conditions in Theorem~\ref{th:optimality_condition} or Corollary~\ref{cl:optimality_condition}. Typically, in such cases, a sequence of approximations is constructed until one of the candidate solutions is sufficiently close to the true optimum. In order to quantify the deviation from the optimum, it is useful to have an upper bound on the optimality gap, i.e., a bound on the difference between the value of the objective function at the candidate solution and at the exact solution. Such a bound is given in the next theorem.
\begin{theorem}[Optimality Gap]
  Let $\bm{q} = (q_1,\ldots,q_N)$ be a solution of \eqref{prob:min_explicit}. For every feasible vector of densities $\bm{p} = (p_1,\ldots,p_N)$ and every $\bm{c} = (c_1, \ldots, c_N) \in \mathbb{R}^N$ it holds that
  \begin{equation*}
    I_f(\bm{p}) - I_f(\bm{q}) \leq \langle \bm{e} , \bm{1} \rangle 
  \end{equation*}
  where $\bm{e} = (e_1, \ldots, e_N)$, with $e_n = e''_n + e'_n$ and
  \begin{align}
    e''_n &= \int_{\Omega} (p_n - \pmax_n) (f_n(\bm{p})-c_n)^- \dint \mu \geq 0, 
    \label{eq:residual_max} \\
    e'_n  &= \int_{\Omega} (p_n - \pmin_n) (f_n(\bm{p}) - c_n)^+ \dint \mu \geq 0. 
    \label{eq:residual_min}
  \end{align}
  \label{th:optimality_gap}
\end{theorem}

A proof of Theorem~\ref{th:optimality_gap} is detailed in Appendix~\ref{apx:proof_of_optimality_gap}. The bound in Corollary~\ref{th:optimality_gap} can be interpreted as a sum of $N$ residuals, where each residual $e_n$ is a measure for how much the density $p_n$ violates the complementary slackness constraint \eqref{eq:compl_slackness}. In turn, $\langle \bm{e} , \bm{1} \rangle = 0$ implies that the optimality conditions in Theorem~\ref{th:optimality_condition} are satisfied and that $\bm{p}$ solves \eqref{prob:min_explicit}. 

In the next section, two numerical algorithms are presented that use the optimality conditions and the bound on the optimality gap to construct a sequence of increasingly accurate approximations to the solution of \eqref{prob:min_explicit}.

\section{Numerical Calculation of the Optimal Densities}
\label{sec:calculation}

For functions $f$ with analytically invertible subderivatives and simple bounds $\pmax_n$, $\pmin_n$, it can be possible to solve the optimality conditions in Theorem~\ref{th:optimality_condition} and Corollary~\ref{cl:optimality_condition} analytically. However, for more complicated problems, the optimality conditions need to be solved numerically. For this case, two efficient and numerically stable algorithms to iteratively approximate the solution of \eqref{prob:min_explicit} are detailed. Both algorithms are based on fixed-point iterations that are solved via a block coordinate descent strategy. The proposed coordinate selection rule is based on the residuals in Theorem~\ref{th:optimality_gap}. While the first algorithm is more efficient and conceptually simpler, it is not guaranteed to converge for every feasible choice of $f$. This shortcoming is overcome in the second algorithm, where an additional outer proximal iteration is introduced to ensure convergence. Both algorithms use the bound on the optimality gap in Corollary~\ref{th:optimality_gap} to track how closely the current iterate approximates the true minimum so that the iteration can be terminated when the desired level of accuracy is achieved.

\subsection{Discretization of the Continuous Problem}

In order to make \eqref{prob:min_explicit} tractable for numerical optimization techniques, it first needs to be reduced to a finite-dimensional problem. A common approach to represent a continuous function $g \colon \Omega \to \mathbb{R}$ by a finite-dimensional vector $\bm{a} \in \mathbb{R}^K$, $K \geq 1$, is to express it in terms of a linear combination of basis functions \cite{EoM_Approximation_of_Functions}, i.e.,
\begin{equation}
  g(\omega) \approx g_{\bm{a}}(\omega) \coloneqq \sum_{k=1}^K a_k \psi_k(\omega-\omega_k),
\label{eq:basis_expansion}
\end{equation}
where $\psi_1,\ldots \psi_K \colon \Omega \to \mathbb{R}$ denote $K$ basis functions centered at grid points $\bm{\omega} = (\omega_1,\ldots,\omega_K)$ and $\bm{a} = (a_1, \ldots, a_K)$ denote the combination weights. 

In an optimization context, the basis functions are typically chosen \emph{a priori}, while the weights are subject to the optimization. For the problem considered in this paper, all basis functions need to be chosen such that they are $\mu$-integrable, i.e.,
\begin{equation*}
  \mu_k \coloneqq \int_{\Omega} \psi_k(\omega) \dint \mu(\omega) < \infty, \quad k = 1, \ldots, K.
\end{equation*}
The masses of the basis functions are collected in a vector $\bm{\mu} = (\mu_1, \ldots, \mu_K)$. Moreover, it is useful to require the basis functions to satisfy
\begin{equation}
  \psi_k(\omega_l-\omega_k) = \begin{dcases}
                                1, & l = k \\
                                0, & l \neq k
                              \end{dcases}
  \label{eq:basis_functions_condition}
\end{equation}
for all $k = 1, \ldots, K$. This assumption is relatively mild and is satisfied by many common sets of basis functions, in particular by appropriately chosen M-splines and squared sinc functions. It guarantees that on the grid points the approximated function $g_{\bm{a}}$ evaluates to the corresponding combination weight, i.e.,
\begin{equation*}
  g_{\bm{a}}(\omega_k) = a_k, \quad k = 1, \ldots, K.
\end{equation*}
This property decouples the combination weights and enables a parallel implementation of the algorithms presented in the following sections. Note that \eqref{eq:basis_functions_condition} does not imply orthogonality.

\subsection{Discrete Optimality Conditions}

In order to obtain a discrete version of the optimality conditions in Theorem~\ref{th:optimality_condition} and Corollary~\ref{cl:optimality_condition}, all densities $p_1, \ldots, p_N$ are again expressed as combinations of a fixed set of basis functions, i.e.,
\begin{equation}
  p_n(\omega) \approx p_{\bm{a}_n}(\omega) \coloneqq \sum_{k=1}^K a_{n,k} \psi_k(\omega-\omega_k), 
  \quad n = 1, \ldots, N,
  \label{eq:basis_expansion_p}
\end{equation}
where $\bm{a}_n \in \mathbb{R}^K$. To facilitate compact notation, the row vectors $\bm{a}_n$ are stacked into a matrix $\bm{A} \in \mathbb{R}^{N \times K}$. The minimization in \eqref{prob:min_implicit} can then be written as the finite-dimensional convex optimization problem
\begin{gather}
  \min_{\bm{A} \in \mathbb{R}^{N \times K}} \; \int_{\Omega} f(\omega, p_{\bm{a}_1}(\omega), \ldots, p_{\bm{a}_N}(\omega) ) \dint\mu
  \label{eq:min_problem_finite}\\[0.5ex]
  \text{s.t.} \quad \pmin_n(\bm{\omega}) \leq \bm{a}_n \leq \pmax_n(\bm{\omega}), \quad \langle\bm{a}_n, \bm{\mu} \rangle = 1, \quad n = 1, \ldots, N. \notag
\end{gather}
Since the degrees of freedom in \eqref{eq:min_problem_finite} are reduced to $NK$, the constraints on each density $p_{\bm{a}_n}$ can only be satisfied at the $K$ grid points. Accordingly, the optimality conditions for problem \eqref{eq:min_problem_finite} are obtained from the general conditions in Theorem~\ref{th:optimality_condition} or Corollary~\ref{cl:optimality_condition} by evaluating them on the grid $\bm{\omega}$. In particular, the optimality condition in Corollary~\ref{cl:optimality_condition} becomes
\begin{equation}
  \bm{a}_n = \min\{\, \pmax_n(\bm{\omega}) \,,\, \max\{\, f_n^{-1}(\remidx{A}{n}, c_n) \,,\, \pmin_n(\bm{\omega}) \,\} \},
  \label{eq:optimality_condition_discrete}
\end{equation}
for all $n = 1, \ldots, N$. In analogy to $f_n^{-1}(\remidx{q}{n},c_n)$, the inverse function $f_n^{-1}(\remidx{A}{n},c_n)$ in \eqref{eq:optimality_condition_discrete} is defined  as
\begin{equation}
  f_n^{-1}(\remidx{A}{n},c_n) = \inf \{ \bm{a}_n \in \mathbb{R}^K : f_n(\bm{A}) \geq c_n \bm{1} \},
  \label{eq:inverse_function_finite}
\end{equation}
where the infimum is taken element-wise and $f_n(\bm{A})$ is shorthand for the vector $f_n(\bm{a}_1, \ldots, \bm{a}_N)$.

\subsection{Discrete Optimality Gap}

Given a feasible vector of densities $\bm{p}$, the integrals in \eqref{eq:residual_max} and \eqref{eq:residual_min} can in principle be evaluated using standard numerical integration techniques. However, this might be problematic in practice. First, depending on the dimensions of $\Omega$ and the chosen technique, numerical integration can be computationally costly. Second, when solving the discrete problem \eqref{eq:min_problem_finite}, the set of feasible solutions is restricted to densities of the form $p_{\bm{a}_n}$ in \eqref{eq:basis_expansion_p}, which will typically not solve the continuous problem \eqref{prob:min_explicit} exactly. Hence, even if the matrix $\bm{A}$ satisfies the discrete optimality conditions in \eqref{eq:optimality_condition_discrete}, the optimality gap can be nonzero. This property is undesirable since in order to provide a useful termination criterion the optimality gap should converge to zero. That is, it should reflect how accurately a candidate solution solves the discrete problem \eqref{eq:min_problem_finite}, instead of the continuous version \eqref{prob:min_explicit}.

In order to avoid these two problems, the following discrete versions of the residuals in \eqref{eq:residual_max} and \eqref{eq:residual_min} are proposed:
\begin{align}
  \tilde{e}''_n &\coloneqq \langle\, (\bm{a}_n - \pmax_n(\bm{\omega})) (f_n(\bm{A}) - c_n \bm{1})^- \,,\, \bm{\mu} \,\rangle, 
  \label{eq:residual_max_discrete} \\
  \tilde{e}'_n  &\coloneqq \langle\, (\bm{a}_n - \pmin_n(\bm{\omega})) (f_n(\bm{A}) - c_n \bm{1})^+ \,,\, \bm{\mu} \,\rangle.
  \label{eq:residual_min_discrete}
\end{align}
The expressions in \eqref{eq:residual_max_discrete} and \eqref{eq:residual_min_discrete} are obtained from \eqref{eq:residual_max} and \eqref{eq:residual_min} by approximating all continuous functions of $\omega$, i.e., $\pmin_n(\omega), \pmax_n(\omega)$, and $f_n(\bm{p}(\omega))$, via linear combinations of the basis functions $\psi_1, \ldots, \psi_K$ and assuming all $p_n$ to be of the form \eqref{eq:basis_expansion_p}. Hence, \eqref{eq:residual_max_discrete} and \eqref{eq:residual_min_discrete} are approximations of the true residuals whose accuracy depends on how well the chosen basis functions represent the true functions. Being an inner product of two vectors, \eqref{eq:residual_max_discrete} and \eqref{eq:residual_min_discrete} can be evaluated efficiently, even for large grid sizes $K$. Moreover, by construction, the discrete residuals become zero whenever a density satisfies the discrete optimality conditions in 
\eqref{eq:optimality_condition_discrete}. 

\subsection{Block Coordinate Descent}

The first proposed algorithm for numerically approximating the optimal densities is detailed below. It is based on the observation that if $N-1$ optimal densities are given, then the remaining one can be determined by a simple line search over the scalar $c_n$. 

\begin{algorithm}[H]
  \small
  \begin{algorithmic}[1]
    \NumTabs{10}
    \STATE \textbf{input} \par
    \hskip\algorithmicindent partial subderivatives \tab $(f_1,\ldots,f_N)$ \par
    \hskip\algorithmicindent density bounds \tab\tab $(\pmin_1,\ldots,\pmin_N)$ and $(\pmax_1,\ldots,\pmax_N)$ \par
    \hskip\algorithmicindent grid points \tab\tab $\bm{\omega} = (\omega_1, \ldots, \omega_K)$  \par
    \hskip\algorithmicindent tolerance \tab\tab\tab $\varepsilon > 0$
    \STATE \textbf{initialize} \par
    \hskip\algorithmicindent Choose a matrix $\bm{A} \in \mathbb{R}^{N \times K}$ whose rows $\bm{a}_1,\ldots,\bm{a}_N$ satisfy the constraints in (15) and a vector $\bm{c} \in \mathbb{R}^N$.  
    \REPEAT
    \STATE Calculate residuals $\tilde{e}_1, \ldots, \tilde{e}_N$ according to \eqref{eq:residual_max_discrete} and \eqref{eq:residual_min_discrete}.
    \STATE Get index $n^*$ of the largest residual
           \begin{equation*}
             n^* \in \argmax_{n = 1,\ldots,N} \, \{ \tilde{e}_n \}.
           \end{equation*}
    \STATE Find a scalar $c$ such that
          \begin{equation*}
            \langle\, \min\{\, \pmax_{n^*}(\bm{\omega}) \,,\, \max\{\, 
            f_{n^*}^{-1}(\remidx{A}{n^*}, c) 
            \,,\, \pmin_{n^*}(\bm{\omega}) \,\} \} \,,\, \bm{\mu} \,\rangle = 1.
          \end{equation*}
   \STATE Set $c_{n^*} \leftarrow c$ and
          \begin{equation*}
            \bm{a}_{n^*} \leftarrow \min\{\, \pmax_{n^*}(\bm{\omega}) \,,\, \max\{\, 
            f_{n^*}^{-1}(\remidx{A}{n^*}, c_{n^*}) 
            \,,\, \pmin_{n^*}(\bm{\omega}) \,\} \}.
          \end{equation*}
    \UNTIL{$\langle \tilde{\bm{e}} , \bm{1} \rangle \leq \varepsilon$} 
    \RETURN $(\bm{A}, \bm{c},  \tilde{\bm{e}})$
  \end{algorithmic}
  \caption{Block coordinate descent}
  \label{alg:1}
\end{algorithm}

Algorithm~\ref{alg:1} implements a block coordinate descent (BCD) \cite{Xu2013,Wright2015} in the vectors $\bm{a}_1,\ldots,\bm{a}_N$ with a custom coordinate selection rule. It is composed of two main steps, namely:
\begin{enumerate}
  \item the coordinate selection step: calculate the residuals $\tilde{e}_1, \ldots, \tilde{e}_N$ and pick the index $n^*$ corresponding to the largest residual;
  \item the coordinate descent step: determine the optimal vector $\bm{a}_{n^*}$ while keeping the remaining rows of $\bm{A}$ fixed.
\end{enumerate}
The idea behind the coordinate selection step is to optimize the density that has the largest residual and hence the largest contribution to the optimality gap. Although this selection scheme is a heuristic, it provides a well-motivated, systematic rule to choose the next coordinate. See the example in Section~\ref{sec:examples} for a comparison of the proposed rule to a cyclic and a random selection rule. 

In the coordinate descend step of Algorithm~\ref{alg:1}, the weight vector $\bm{a}_n$ is chosen such that $p_{\bm{a}_n}$ satisfies the discrete optimality condition in \eqref{eq:optimality_condition_discrete} and, therefore, is a coordinate-wise minimum of \eqref{eq:min_problem_finite}. The central task in the coordinate descend step is the evaluation of $f_n^{-1}(\remidx{A}{n},c)$. In case the inverse of $f_n$ has an analytical from, the corresponding explicit function can simply be substituted for $f_n^{-1}$. An example for this case is given in Section~\ref{ssec:weighted_kl_div}. Otherwise, $f_n^{-1}(\remidx{A}{n},c)$ needs to be evaluated numerically by solving \eqref{eq:inverse_function_finite} for $\bm{a}_n$. Owing to property \eqref{eq:basis_functions_condition} of the basis functions, the elements of the vector $\bm{a}_n$ in \eqref{eq:inverse_function_finite} can be determined by solving 
\begin{equation}
  f_n(a_{1,k}, \ldots, a_{n,k}, \ldots, a_{N,k}) = c
  \label{eq:root_finding}
\end{equation}
for each $k = 1, \ldots, K$ individually. This decoupling makes it possible to evaluate $f_n^{-1}(\remidx{A}{n},c)$ in a highly parallel manner, using up to $K$ compute cores simultaneously. In addition, the memory requirements per core are minimal since apart from $f_n$ and $c$ only the vector $(a_{1,k}, \ldots,a_{N,k})$ needs to be stored, which is of dimension $N$, irrespective of how $K$ is chosen. 

Since $f_n$ is monotonically nondecreasing for all $n$, \eqref{eq:root_finding} can be solved via standard one-dimensional root-finding techniques. In order to obtain $\bm{a}_n$, the vector $f_n^{-1}(\remidx{A}{n},c)$ is then projected onto the density band $\Pband_n$. Consequently, no root-finding needs to be performed if the root is guaranteed to lie outside the feasible interval $[\pmin_n(\omega_k), \pmax_n(\omega_k)]$. Whether or not this is the case can be determined by simply evaluating $f_n$ at the endpoints of the interval. From
\begin{equation*}
  f_n(a_{1,k}, \ldots, a_{n-1,k}, \pmin_n(\omega_k), a_{n+1,k}, \ldots, a_{N,k}) \geq c
\end{equation*}
it follows that $a_{n,k} < \pmin_n(\omega_k)$ so that, after the projection onto the feasible band, $a_{n,k} = \pmin_n(\omega_k)$. Analogously, it follows from
\begin{equation*}
  f_n(a_{1,k}, \ldots, a_{n-1,k}, \pmax_n(\omega_k), a_{n+1,k}, \ldots, a_{N,k}) \leq c
\end{equation*}
that $a_{n,k} = \pmax_n(\omega_k)$. These additional checks are consistent with the definition of the generalized inverse in \eqref{eq:f_inverse} and simplify the evaluation of $f_n^{-1}(\remidx{A}{n},c)$ significantly.

Algorithm~\ref{alg:1} terminates if the optimality gap becomes smaller than the tolerance $\varepsilon$. It should be emphazised again that $\tilde{\bm{e}}$ in line 8 corresponds to the optimality gap of the discrete problem \eqref{eq:min_problem_finite} so that $\langle \tilde{\bm{e}}, \bm{1} \rangle \leq \varepsilon$ does \emph{not} guarantee $\langle \bm{e}, \bm{1} \rangle \leq \varepsilon$. However, for reasonably small grid sizes both values are sufficiently close. Moreover, in case of strict optimality requirements, the equations in Theorem~\ref{th:optimality_gap} can be used to obtain a tighter bound on the true optimality gap.

Algorithm~\ref{alg:1} is a straightforward attempt at solving the system of optimality conditions in Theorem~\ref{th:optimality_condition}. By construction, its limit points satisfy \eqref{eq:optimality_condition_discrete} and, hence, are global minimizers of \eqref{eq:min_problem_finite}. However, for general convex functions $f$, its convergence cannot be guaranteed since the block coordinate descent strategy implemented in Algorithm~\ref{alg:1} is not guaranteed to converge for objective functions that are not strictly convex \cite{Tseng2001}. This problem is intrinsic to coordinate-wise minimization techniques and exists irrespective of the coordinate selection rule and the termination criterion. A proximal algorithm that addresses these shortcomings at the cost of a reduced efficiency is detailed in the next section.

\subsection{Proximal BCD With Guaranteed Convergence}

The algorithm presented in this section solves the system of optimality conditions by means of a proximal iteration instead of a regular fixed-point iteration. It extends Algorithm~\ref{alg:1} to cover cases where it might otherwise fail to converge.

Proximal algorithms are well studied and offer a reliable tool for iteratively solving optimization problems that are not strictly convex. The underlying idea is to construct a sequence of strictly convex objective functions whose minimizers converge to the minimizer of the original problem. To this end, the original objective function is augmented with a strictly convex term such that the solution of the augmented problem is guaranteed to be unique. More precisely, in each iteration, a proximal algorithm seeks to minimize a weighted sum of the objective function and a term that penalizes some distance between the current and the previous iterate. The additional distance term ensures that the problem is strictly convex and automatically vanishes when a minimum of the orginal objective function is approached. A comprehensive introduction to proximal algorithms can be found in \cite{Parikh2014}.

For the function $I_f$ in \eqref{def:functional}, a proximal operator based on the $\mathcal{L}_{\mu}^2$-norm can be defined as
\begin{align}
  &\prox{I_f}{h_1,\ldots,h_n} \notag \\
  &\coloneqq \argmin_{ \{p_n \in \Pband_n\}_{n=1}^N } \; \left( I_f(p_1,\ldots,p_N) + \frac{1}{2} \sum_{n=1}^N  \lVert p_n - h_n \rVert_2^2 \right) \notag \\
  &= \argmin_{ \{p_n \in \Pband_n\}_{n=1}^N} \; \left( \int_{\Omega} f(\bm{p}) + \frac{1}{2} \sum_{n=1}^N  (p_n - h_n)^2 \dint \mu \right),
  \label{def:proximal_operator}
\end{align}
where $h_1, \ldots, h_N \colon \Omega \to \mathbb{R}$ are $\mu$-integrable functions. The proximal algorithm iteratively approximates the solution of the fixed-point equation
\begin{equation*}
  (h_1,\ldots,h_N) = \prox{I_f}{h_1,\ldots,h_n}.
\end{equation*}
In order to implement an iteration of the proximal algorithm, the inner problem in \eqref{def:proximal_operator} needs to be solved. By inspection, this problem is equivalent to the original problem \eqref{prob:min_explicit} with $f$ replaced by
\begin{equation*}
  \tilde{f}(\omega,\bm{x}) \coloneqq f(\omega,\bm{x}) + \frac{1}{2} \sum_{n=1}^N (x_n - h_n(\omega))^2,
\end{equation*}
which is strictly convex in $\bm{x}$ for every convex function $f$ and admits the partial subderivatives
\begin{equation}
  \tilde{f}_n(\omega,\bm{x}) = f_n(\omega,\bm{x}) + x_n - h_n(\omega).
  \label{eq:f_proximal_derivative}
\end{equation}
The proximal block coordinate descent algorithm is denoted Algorithm~\ref{alg:2} and is specified below.

\begin{proposition}[Convergence of Algorithm~\ref{alg:2}]
  Let the basis functions in \eqref{eq:basis_expansion} be chosen such that $\bm{\mu} > 0$. For all convex functions $f$ and all bands $\Pband$ that satisfy
  \begin{equation*}
  \langle \pmax(\bm{\omega}), \bm{\mu} \rangle \geq 1 \quad \text{and} \quad
  \langle \pmin(\bm{\omega}), \bm{\mu} \rangle \leq 1,
  \end{equation*}
  Algorithm~\ref{alg:2} is guaranteed to converge to a global minimizer of \eqref{eq:min_problem_finite}.
  \label{prop:convergence}
\end{proposition}

A proof for Proposition~\ref{prop:convergence} can be found in Appendix~\ref{apx:proof_of_convergence_algorithm_2}. The two conditions on the density bands are vectorized versions of the integral inequalities in \eqref{eq:band_conditions} and ensure that in each iteration a feasible density of the form \eqref{eq:basis_expansion} can be constructed.

\begin{algorithm}[H]
  \small
  \begin{algorithmic}[1]
    \NumTabs{10}
    \STATE \textbf{input} \par
    \hskip\algorithmicindent partial subderivatives \tab $(f_1,\ldots,f_N)$ \par
    \hskip\algorithmicindent density bounds \tab\tab $(\pmin_1,\ldots,\pmin_N)$ and $(\pmax_1,\ldots,\pmax_N)$ \par
    \hskip\algorithmicindent grid points \tab\tab $\bm{\omega} = (\omega_1, \ldots, \omega_K)$  \par
    \hskip\algorithmicindent tolerance \tab\tab\tab $\varepsilon > 0$
    \STATE \textbf{initialize} \par
    \hskip\algorithmicindent Choose a matrix $\bm{A} \in \mathbb{R}^{N \times K}$ whose rows $\bm{a}_1,\ldots,\bm{a}_N$ satisfy the constraints in (15) and a vector $\bm{c} \in \mathbb{R}^N$.  
    \REPEAT
      \STATE For all $n = 1,\ldots,N$ set
             \begin{equation*}
               \tilde{f}_n(\omega,x_1,\ldots,x_N) \leftarrow f_n(\omega,x_1,\ldots,x_N) + x_n 
                  - p_{\bm{a}_n}\!(\omega).
             \end{equation*}
      \STATE Set
             \begin{equation*}
               \textbf{input} \leftarrow \left\{ (\tilde{f}_1,\ldots,\tilde{f}_N),
                (\pmin_1,\ldots,\pmin_N), (\pmax_1,\ldots,\pmax_N), \bm{\omega}, \varepsilon 
                \right\}.
             \end{equation*}
      \STATE Use Algorithm~\ref{alg:1} to update $\bm{A}$ and $\bm{c}$
             \begin{equation*}
               (\bm{A}, \bm{c}) \leftarrow \textbf{Algorithm~\ref{alg:1}} (\textbf{input}) 
             \end{equation*}
      \STATE Calculate residuals $\tilde{e}_1, \ldots, \tilde{e}_N$ according to 
             \eqref{eq:residual_max_discrete} and \eqref{eq:residual_min_discrete} using the 
             original derivatives $(f_1, \ldots, f_N)$.
    \UNTIL $\langle \tilde{\bm{e}} , \bm{1} \rangle < \varepsilon$
    \RETURN $(\bm{A}, \bm{c},  \tilde{\bm{e}})$
  \end{algorithmic}
  \caption{Proximal BCD with guaranteed convergence}
  \label{alg:2}
\end{algorithm}

The price for the guaranteed convergence of the proximal version of the block coordinate descent algorithm is reduced efficiency. Since Algorithm~\ref{alg:2} repeatedly calls Algorithm~\ref{alg:1} to solve the minimization in \eqref{def:proximal_operator}, it requires as least as many iterations as  Algorithm~\ref{alg:1}, assuming that the latter converges. The total number of coordinate descent steps of Algorithm~\ref{alg:2} is roughly given by the product of the number of iterations required by both algorithms. Moreover, in some cases an analytic expression for the inverse of $f_n$ exists, but not for the inverse of $\tilde{f}_n$. In general, the use of Algorithm~\ref{alg:2} is recommended only when Algorithm~\ref{alg:1} indeed fails to converge.

\subsection{Remarks}

There are several options to improve the performance of the presented algorithms. First, if evaluating the residuals is significantly more expensive than the coordinate descent step, the coordinate selection rule can be modified to only evaluate the residuals every $M$th $(M > 1)$ iteration or a simple cyclic or random selection rule can be used \cite{Saha2013,Nutini2015}. Second, in Algorithm~\ref{alg:2}, the weight of the penalty term can be reduced in order to reduce the number of outer iterations. In practice, this means trading off speed of convergence for numerical stability. 

Apart from the two algorithms presented in this section, the finite-dimensional problem \eqref{eq:min_problem_finite} can, in principle, be solved by many off-the-shelf convex optimization programs. The proposed algorithms are often preferable over generic solvers for the following reasons:

First, the approach presented in this paper often allows an analytical general form for the optimal densities to be obtained. The numerical part of the solution process then reduces to determining the scalars $c_1,\ldots,c_N$, which usually is a much simpler problem. Moreover, having a parametric expression for the least favorable densities can facilitate the derivation of analytical bounds or approximations to the exact solution. An example where the system of optimality conditions leads to a useful lower bound is given in the next section.

Second, the accuracy of generic convex optimization algorithms can be hard to control for certain density bands. If, for example, in some region of the sample space the width of the band is of the same order of magnitude as the tolerance of the solver, all points within the band become equivalent. While this effect is negligible in terms of the shape of the optimal densities themselves, it can be critical when calculating functions that involve products or ratios of optimal densities. Although the amount of numerical noise in the solution can be reduced by applying suitable variable transformations and carefully tuning the absolute and relative tolerances of the solvers, the general issue of having to deal with, potentially, very badly scaled problems is an important consideration.

The presented algorithms avoid scaling problems by reducing the entire optimization process to repeated searches for the root of a nondecreasing real-valued function. In each iteration, both $c_n$ and, if necessary, $\bm{a}_{n}$ are determined via one-dimensional root-finding. The latter is a standard problem in numerical mathematics and can be solved fast and reliably---see, for example, \cite[Chapter 8.3]{Monahan2001} for an overview of suitable algorithms. In cases where $f_n$ can only be evaluated with (numerical) noise, stochastic root-finding methods, which handle the additional uncertainty in a systematic manner \cite{Waeber2013}, can be applied. Owing to the explicit projection on $\Pband_n$, the optimal densities are exact in regions of $\Omega$ where the band constraints are active.

The third advantage of the proposed algorithms is that they can be significantly faster than generic solvers. The most substantial performance gains can be achieved with Algorithm~\ref{alg:1} if an analytic expression for the inverse of $f_n$ exists; see the example in the next section. However, even if $f_n$ has to be inverted numerically, this can be done in a highly parallel manner that facilitates the use of high performance graphics processing units and distributed computing systems. Note that neither of the presented algorithms needs to perform matrix operation (multiplication, inversion) which makes them highly scalable in terms of the number of densities $N$ and the number of grid points $K$. This is illustrated with an example in the next section.

In summary, the proposed approach to the minimization of convex functionals offers useful theoretical insights and at the same time provides a fast and reliable way to obtain accurate numerical results.

\section{Examples}
\label{sec:examples}

In this section, the usefulness of the results presented in the previous sections is demonstrated by means of two examples. First, the convex functional is chosen as a weighted sum of Kullback--Leibler divergences. This example shows how analytical results can be obtained by means of Theorem~\ref{th:optimality_condition} and it demonstrates the increased efficiency and accuracy of the proposed algorithms in comparison to a state-of-the-art generic solver. In the second example, the least favorable distributions for a binary decision making problem with an observation dependent cost function are derived. This example illustrates how the concept of minimizing convex functionals is applicable beyond its traditional context of statistical distance measures.

In order to simplify the presentation, the sample space is chosen to be the real line, i.e., $\Omega = \mathbb{R}$. For the finite-dimensional representation of the densities, a regular grid with step size $\omega_k - \omega_{k-1} = \Delta \omega$ is used in combination with a linear interpolation scheme, i.e.,
\begin{equation*}
  \psi_k(\omega-\omega_k) = \begin{dcases}
                              1-\frac{\lvert \omega-\omega_k \rvert}{\Delta \omega}, & \omega \in [\omega_{k-1}, \omega_{k+1}] \\
                              0, & \text{otherwise}
                            \end{dcases}.
\end{equation*}
The masses of the basis functions calculate to $\mu_k = \Delta \omega$ for all $k = 1, \ldots, K$ and a simple bisection algorithm was used to perform the root finding.

\subsection{Weighted Sum of Kullback--Leibler Divergences}
\label{ssec:weighted_kl_div}

Sums of $f$-divergences, and Kullback--Leibler divergences in particular, have been shown to have applications in a variety of fields, including minimax robust statistics \cite{Guntuboyina2011}, geoscience \cite{Karoui2009} and biology \cite{Suzuki2004}. In this example, $I_f$ is a weighted sum of Kullback--Leibler divergences with respect to a common reference distribution $P_N$, i.e.,
\begin{equation}
  I_f(p_1,\ldots,p_N) = \sum_{n=1}^{N-1} \alpha_n D_{\text{KL}}(p_N \,\Vert\, p_n),
  \label{eq:functional_weighted_kl}
\end{equation}
where $D_{\text{KL}}(\cdot \Vert \cdot)$ denotes the Kullback-Leibler divergence and $\alpha_1,\ldots,\alpha_{N-1}$ are convex combination weights, i.e., they satisfy
\begin{equation*}
   \alpha_1, \ldots, \alpha_{N-1} \geq 0, \qquad \sum_{n=1}^{N-1} \alpha_n = 1.
\end{equation*}
The corresponding function $f$ is given by
\begin{equation}
  f(\omega,x_1,\ldots,x_N) = \sum_{n=1}^{N-1} \alpha_n \log \biggl( \frac{x_N}{x_n} \biggr) x_N.
  \label{eq:function_weighted_kl}
\end{equation}
Its partial derivatives are
\begin{equation*}
  f_n(\omega,x_1,\ldots,x_N) = -\alpha_n\frac{x_N}{x_n}
\end{equation*}
for $n = 1, \ldots, N-1$ and
\begin{equation*}
  f_N(\omega,x_1,\ldots,x_N) = 1 + \sum_{n=1}^{N-1} \alpha_n \log \biggl( \frac{x_N}{x_n} \biggr).
\end{equation*}
The inverse functions are obtained by solving $f_n = c_n$ for $x_n$ and are given by
\begin{equation*}
  f_n^{-1}(\remidx{x}{n},c_n) = -\frac{\alpha_n}{c_n} x_N \eqqcolon b_n x_N
\end{equation*}
for $n = 1, \ldots, N-1$ and
\begin{equation*}
  f_N^{-1}(\remidx{x}{N},c_n) = e^{c_N-1} \prod_{n=1}^{N-1} x_n^{\alpha_n} \eqqcolon b_N \prod_{n=1}^{N-1} x_n^{\alpha_n}.
\end{equation*}
Here $b_1,\ldots,b_{N-1} \in \mathbb{R}$ and $b_N > 0$ are introduced for the sake of a more compact notation. From Corollary~\ref{cl:optimality_condition} it follows that the optimal densities are of the form
\begin{equation*}
  q_n = \min \{\, \pmax_n \,,\, \max \{\, b_n q_N \,,\, \pmin_n \,\} \}
\end{equation*}
for $n = 1, \ldots, N-1$ and
\begin{equation}
  q_N = \min \{\, \pmax_N \,,\, \max \{\, b_N q_1^{\alpha_1} \cdots q_{N-1}^{\alpha_{N-1}} \,,\, \pmin_N \,\} \}.
  \label{eq:optimal_density_N}
\end{equation}
That is, $q_1,\ldots,q_{N-1}$ are the projections of $q_N$ onto the bands $\Pband_1, \ldots, \Pband_{N-1}$, respectively, and $q_N$ is the projection of the weighted geometric mean of $q_1, \ldots, q_{N-1}$ onto the band $\Pband_N$.

Before presenting numerical results, it is shown that the above expressions for the optimal densities can be used to derive a tight lower bound on \eqref{eq:functional_weighted_kl} for the special case where $p_1,\ldots,p_{N-1}$ are given and the optimization is performed only over $p_N$. Problems of this kind are, for example, considered in \cite{Guntuboyina2011} in order to derive lower bounds on the minimax risk of a decision making procedure. Making use of \eqref{eq:optimal_density_N}, it can be shown that
\begin{align*}
  \min_{p_N \in \Pband_N} I_f(p_1,\ldots,p_N) &\geq \min_{p_N \in \mathcal{L}_{\mu}^1} I_f(p_1,\ldots,p_N) \\
  &= \sum_{n=1}^{N-1} \alpha_n D_{\text{KL}}( b_N p_1^{\alpha_1} \cdots p_{N-1}^{\alpha_{N-1}} \,\Vert\, p_n ) \\
  &= -\log b_N,
\end{align*}
where
\begin{equation*}
  b_N = \int_{\Omega} p_1^{\alpha_1} \cdots p_{N-1}^{\alpha_{N-1}} \dint \mu
\end{equation*}
is a generalized version of the Bhattacharyya coefficient \cite{Bhattacharyya1943}. To the best of our knowledge, this bound has not been stated in the literature so far.

For the numerical minimization of \eqref{eq:functional_weighted_kl} consider $N=3$ densities and uncertainty bands defined by scaling and shifting Gaussian densities according to
\begin{equation}
  \begin{aligned}
    \pmin_1 &= 0.8 \, p_{\mathcal{N}}(-0.5,1), &
    \pmax_1 &= 1.2 \, p_{\mathcal{N}}(-0.5,1), \\
    \pmin_2 &= 0.8 \, p_{\mathcal{N}}(\phantom{-}0.5,1), &
    \pmax_2 &= 1.2 \, p_{\mathcal{N}}(\phantom{-}0.5,1), \\
    \pmin_3 &= 0.8 \, p_{\mathcal{N}}(\phantom{-0.}0,1), &
    \pmax_3 &= 1.2 \, p_{\mathcal{N}}(\phantom{-0.}0,1),
  \end{aligned}
  \label{eq:bands_weighted_kl}
\end{equation}
where $p_{\mathcal{N}}(m,\sigma^2)$ denotes the density function of a Gaussian distribution with mean $m$ and variance $\sigma^2$.

\begin{figure}[!t]
  \centering
  \subfloat{\includegraphics{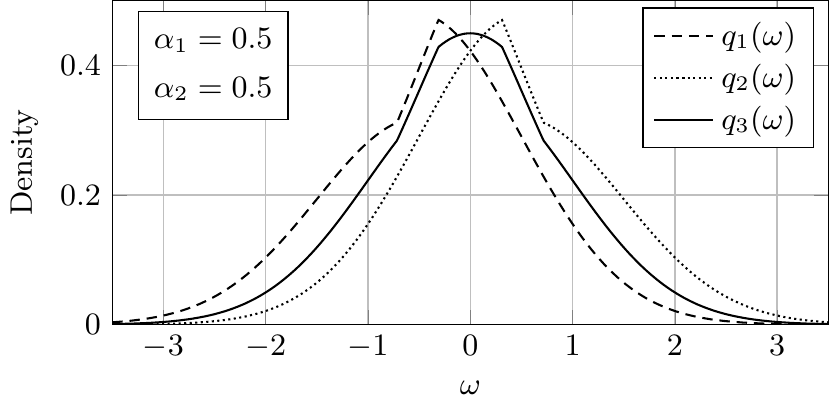} \label{sfig:lfds_wkl_05_05}} \\[0.1cm]
  \subfloat{\includegraphics{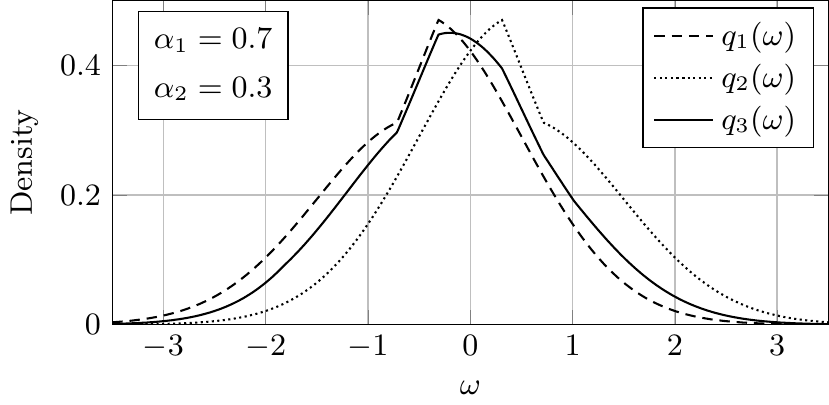} \label{sfig:lfds_wkl_07_03}} \\[0.1cm]
  \subfloat{\includegraphics{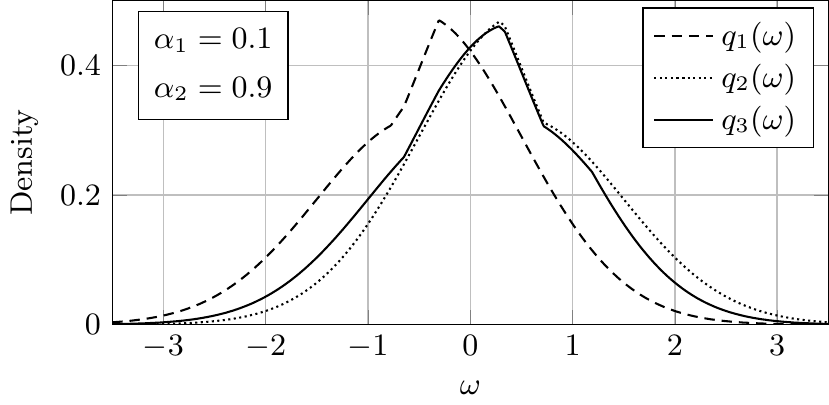} \label{sfig:lfds_wkl_01_09}}
  \caption{Densities that minimize the weighted sum of two Kullback-Leibler divergences for different weights $\alpha_1$ and $\alpha_2$ as calculated by Algorithm~\ref{alg:1}. The objective function is given in \eqref{eq:functional_weighted_kl} ($N = 3$), the density bands in \eqref{eq:bands_weighted_kl}.}
  \label{fig:weighted_kl_densities}
\end{figure}

Three triplets of optimal densities, for different weights $\alpha_1$ and $\alpha_2$, are depicted in Fig.~\ref{fig:weighted_kl_densities}. As can be seen, $q_1$ and $q_2$ are independent of the weights in this particular example, but $q_3$ changes significantly and different combinations push it either towards $q_1$ or $q_2$.

The densities in Fig.~\ref{fig:weighted_kl_densities} were calculated using Algorithm~\ref{alg:1} on the interval $[-5,5]$ with step size $\Delta \omega = 0.01$ and initial densities $q_1 = p_{\mathcal{N}}(-0.5,1)$, $q_2 = p_{\mathcal{N}}(0.5,1)$, $q_3 = p_{\mathcal{N}}(0,1)$. The tolerance for the optimality gap was set to $\varepsilon = 10^{-7}$. The number of iterations required to reach convergence is shown in Table~\ref{tb:iterations} for different coordinate selection rules. As can be seen, a selection based on the largest residual is preferable for asymmetric weights $\alpha_1, \alpha_2$, while for symmetric weights the cyclic rule performs best. As expected \cite{Nutini2015}, random coordinate selection, i.e., drawing the next coordinate from the set $\{1, \ldots, N\} \setminus \{n^*\}$, performs uniformly worst. Although these results are not conclusive, they indicate that a selection based on the largest residual is less sensitive to the shape of the objective function compared to a cyclic rule.

\begin{table}[!t]
  \centering
  \includegraphics{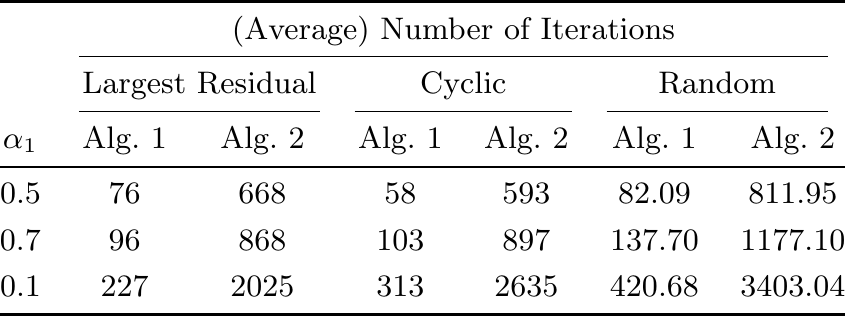}
  \caption{Number of coordinate descent iterations required by the proposed algorithms to minimize the objective \eqref{eq:functional_weighted_kl} under band constraints \eqref{eq:bands_weighted_kl} for different coordinate selection rules and a tolerance of $\varepsilon = 10^{-7}$. The results of the random selection rule are averaged over 100 runs.}
  \label{tb:iterations}
\end{table}

In this example, the inverse functions $f_n^{-1}$ do not need to be evaluated numerically and Algorithm~\ref{alg:1} is highly efficient. Our \textsc{Matlab}\textcopyright{} implementation terminated in well under a second on a regular desktop computer.\footnote{
  All simulations were performed on an Intel\textsuperscript{\textregistered} Core\texttrademark{} i5-760@2.80GHz using \textsc{Matlab}\textcopyright{} 2016a.
}
For comparison, the minimization problem was also solved using version 2.0.4 of the ECOS solver \cite{Domahidi2013}, which is a state-of-the-art software package for solving conic optimization problems and is written in C. The reason for choosing ECOS over other options is that it is one of the few high performance solvers that support the exponential cone and, hence, logarithmic objective functions. The average run-times of Algorithm~\ref{alg:1}, using the largest residual coordinate selection rule, and the ECOS solver are given in Table~\ref{tb:runtime}. The results were obtained on the same machine with absolute tolerances set to $10^{-7}$ for all algorithms. The ECOS solver was called via its CVX interface \cite{cvx}, but only the time spent in the C routine was used for the benchmark. As can be seen, Algorithm~\ref{alg:1} is consistently faster than the ECOS solver, despite the additional handicap of being written in an interpreted language. Especially for large numbers of grid points $K$, the advantage of the proposed approach becomes obvious.

\begin{table}[!t]
  \centering
  \includegraphics{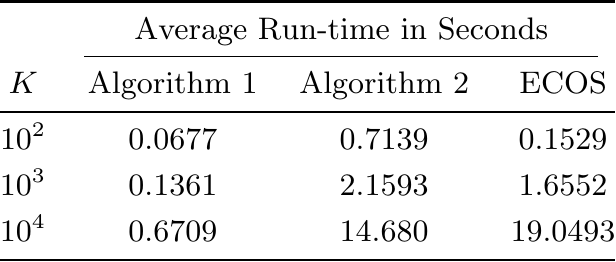}
  \caption{Run-times required by Algorithm~\ref{alg:1}, Algorithm~\ref{alg:2}, and the ECOS solver to minimize objective \eqref{eq:functional_weighted_kl} with weights $\alpha_1 = 0.7$, $\alpha_2 = 0.3$ under band constraints \eqref{eq:bands_weighted_kl} for different grid sizes $K$. The absolute tolerance was set to $\varepsilon = 10^{-7}$ for all algorithms and the results were averaged over 20 runs.}
  \label{tb:runtime}
\end{table}

Table~\ref{tb:iterations} and Table~\ref{tb:runtime} also include the number of iterations and run-times of Algorithm~\ref{alg:2}. It can be shown that for $f$ of the form \eqref{eq:function_weighted_kl} the inverses of the partial derivatives in \eqref{eq:f_proximal_derivative} calculate to
\begin{equation*}
  \tilde{f}_n^{-1}(\remidx{x}{n},c_n) = \frac{c_n+h(\omega)}{2} + \sqrt{ \left( \frac{c_n+h(\omega)}{2} \right)^2 + \alpha_n x_N }
\end{equation*}
for $n \in \{1, \ldots, N\}$ and
\begin{equation*}
  \tilde{f}_N^{-1}(\remidx{x}{N},c_N) = W \Biggl( e^{c_N+h_N(\omega)-1} \prod_{n=1}^{N-1} x_n^{\alpha_n} \Biggr).
\end{equation*}
Here, $W$ denotes the Lambert W function. Table~\ref{tb:iterations} shows the number of inner iterations for Algorithm~\ref{alg:2}, i.e., the total number of coordinate descent steps. The results were obtained by calling Algorithm~1 with the specified selection rule in each iteration of Algorithm~2. The number of outer iterations, i.e., the number of times Algorithm~\ref{alg:1} was called from within Algorithm~\ref{alg:2}, is (approximately) independent of the coordinate selection rule. In this example, Algorithm~\ref{alg:2} required roughly $40$, $50$, and $110$ outer iterations for $\alpha_1 = 0.5$, $\alpha_1 = 0.7$, and $\alpha_1 = 0.1$, respectively. Owing to the increased number of coordinate descent steps, and the computationally costly evaluation of the Lambert W function, Algorithm~\ref{alg:2} is also significantly slower than Algorithm~\ref{alg:1}. However, for medium to large problem sizes its performance is comparable to, or even better than, that of the ECOS solver.

These results arise from the chosen example and it is noted that raw execution time is not a reliable performance metric. Nevertheless, the example is non-trivial and the fact that the proposed algorithms are able to outperform an optimized software package is indicative of a good performance in general.

\begin{figure}[!t]
  \centering
  \includegraphics{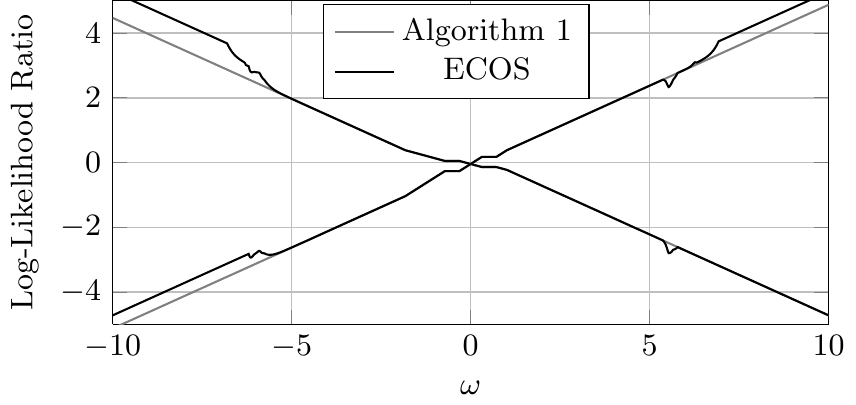}
  \caption{Log-likelihood ratios of the optimal densities shown in the second plot of Fig.~\ref{fig:weighted_kl_densities} as calculated by Algorithm~\ref{alg:1} and the ECOS solver.}
  \label{fig:llrs}
\end{figure}

Another important aspect of the proposed algorithms is that they lead to results with high accuracy. This can be seen by inspection of the ratios of the optimal densities, which are of particular interest in detection problems, where they determine the optimal test statistic. For $\alpha_1 = 0.7$ and $\alpha_2 = 0.3$, the log-likelihood ratios
\begin{equation*}
  \log \frac{q_1(\omega)}{q_3(\omega)} \quad \text{and} \quad \log \frac{q_2(\omega)}{q_3(\omega)},
\end{equation*}
as calculated by the proposed algorithms and the ECOS solver, are depicted in Fig.~\ref{fig:llrs}. Since the objective function in \eqref{eq:functional_weighted_kl} is strictly convex, Algorithm~\ref{alg:1} and Algorithm~\ref{alg:2} converge to the same solution and only the result of Algorithm~\ref{alg:1} is depicted in Fig.~\ref{fig:llrs}. Note that the interval of the sample space is increased to $[-10,10]$. It can clearly be seen how the ECOS solver produces artifacts at the tails of the densities, where their values fall below the tolerance of $10^{-7}$. Algorithm~\ref{alg:1}, in contrast, correctly identifies the tails as regions where the band constraints are active and is, therefore, able to calculate the likelihood ratios exactly. If at all possible, obtaining results of comparable quality with generic solvers requires careful parameter tuning and tolerances close to the machine precision.

\subsection{Minimax Detection with Observation Dependent Cost}

The purpose of the second example is to show that the proposed algorithms handle cases where $f$ directly depends on $\omega$ and that this dependence widens the scope of problems which can be solved within the presented framework.

Consider a binary decision making problem with a cost function
\begin{equation}
  r(\delta) = E_{P_1}\bigl[ \delta(\omega) r_1(\omega) \bigr] + E_{P_2}\bigl[ (1-\delta(\omega))r_2(\omega) \bigr],
  \label{eq:cost_function}
\end{equation}
where $E_P$ denotes expectation with respect to a distribution $P$, $\delta \colon \Omega \to [0,1]$ denotes a decision rule, and $r_0, r_1 \colon \Omega \to [0,\infty)$ denote observation dependent costs for each  decision. Cost functions of this form occur, for example, in detection problems, where the cost for an incorrect decision depends on the true state of the system. In a collision avoidance system for vehicles, for instance, the cost for not detecting an obstacle becomes higher when the obstacle is close to the vehicle or the vehicle is moving at a high speed.

\begin{figure}[!t]
  \centering
  \includegraphics{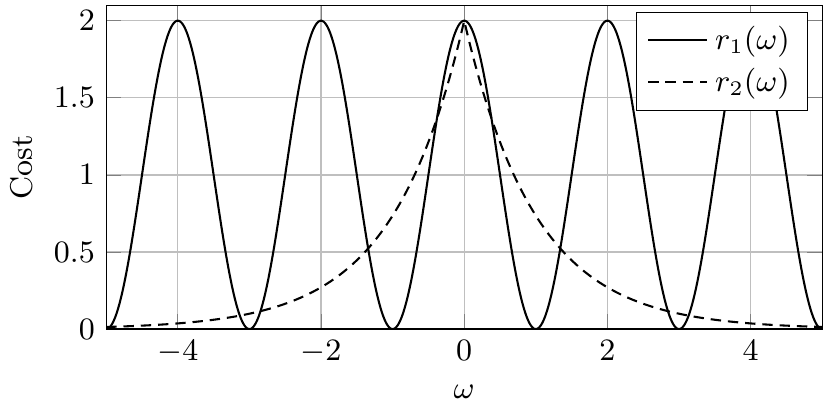}
  \caption{Cost functions $r_1$ and $r_2$ as defined in \eqref{eq:r1_r2}.}
  \label{fig:cost_functions}
\end{figure}

\begin{figure}[!t]
  \centering
  \includegraphics{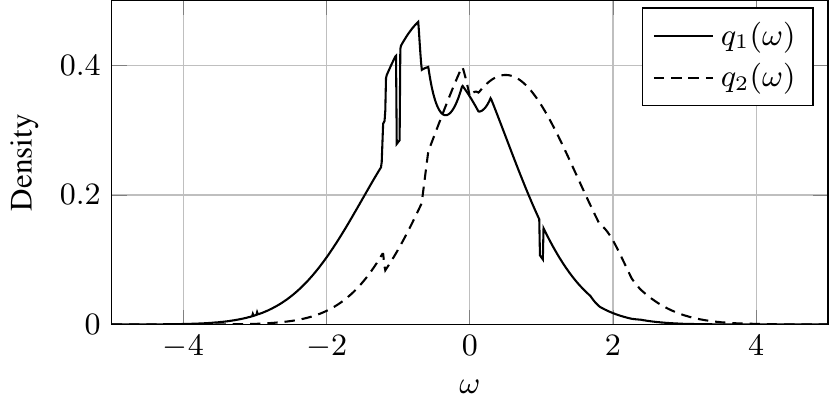}
  \caption{Least-favorable densities as calculated by Algorithm~1 for the cost function \eqref{eq:cost_function} with $r_1$ and $r_2$ chosen according to \eqref{eq:r1_r2} and density bands $\Pband_1, \Pband_2$ according to \eqref{eq:bands_weighted_kl}.}
  \label{fig:cost_min_lfds}
\end{figure}

\begin{figure}[!t]
  \centering
  \includegraphics{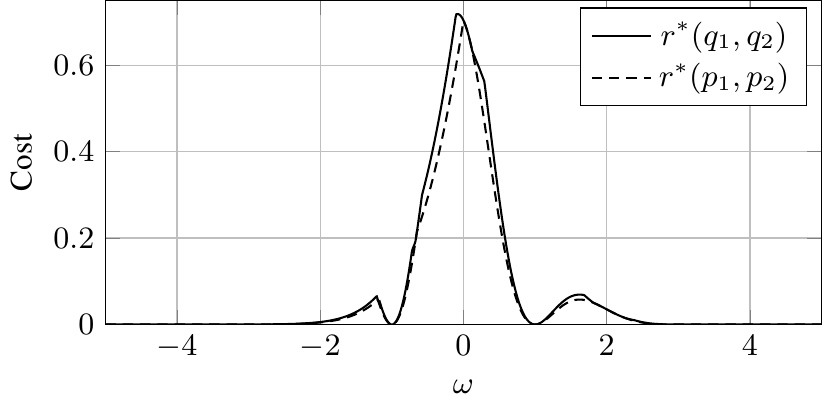}
  \caption{Cost function $r^*$ in \eqref{eq:minimum_cost} for Gaussian densities $p_1 = p_{\mathcal{N}}(-0.5,1)$, $p_2 = p_{\mathcal{N}}(0.5,1)$ and for the least favorable densities $q_1$, $q_2$ in Fig.~\ref{fig:cost_min_lfds}.}
  \label{fig:minimax_cost}
\end{figure}

The optimal decision rule for the cost function \eqref{eq:cost_function} can be shown to be given by 
\begin{equation*}
  \delta^*(\omega) = \begin{dcases}
                       1,                & r_1(\omega)p_1(\omega) < r_2(\omega)p_2(\omega) \\
                       \kappa \in [0,1], & r_1(\omega)p_1(\omega) = r_2(\omega)p_2(\omega) \\
                       0,                & r_1(\omega)p_1(\omega) < r_2(\omega)p_2(\omega)
                     \end{dcases}.
\end{equation*}
Without loss of generality, it is assumed that $\kappa = 1$. Using this decision rule, the expected cost is given by
\begin{equation}
  \int_{\Omega} \min\{ r_1 p_1, r_2 p_2\} \dint \mu \eqqcolon \int_{\Omega} r^*(p_1,p_2) \dint \mu.
  \label{eq:minimum_cost}
\end{equation}
It is assumed that $p_1$ and $p_2$ are subject to uncertainties of the density band type. In order to design a minimax detector, i.e., a detector that minimizes the worst-case cost, \eqref{eq:minimum_cost} needs to be maximized with respect to the densities $p_1$ and $p_2$. This problem is of the form \eqref{prob:min_implicit} with
\begin{equation}
  -f(\omega,x_1,x_2) = \min\{ r_1(\omega)x_1, r_2(\omega)x_2\}
  \label{eq:f_example_2}
\end{equation}
and
\begin{align*}
  -f_1(\omega,x_1,x_2) = \begin{dcases}
                          r_1(\omega), &  r_1(\omega)x_1 \leq r_2(\omega)x_2 \\
                          0, & \text{otherwise} 
                        \end{dcases}, \\
  -f_2(\omega,x_1,x_2) = \begin{dcases}
                          r_2(\omega), &  r_1(\omega)x_1 > r_2(\omega)x_2 \\
                          0, & \text{otherwise}
                        \end{dcases}.
\end{align*}
For illustration purposes, the cost functions are chosen as
\begin{equation}
  r_1(\omega) = 1+\cos(\pi \omega) \quad \text{and} \quad r_2(\omega) = 2\exp(-\lvert \omega \rvert).
  \label{eq:r1_r2}
\end{equation}
Their graphs are shown in Fig.~\ref{fig:cost_functions}. The same density bands as in \eqref{eq:bands_weighted_kl} are used to constrain $p_1$ and $p_2$. The grid for the discrete representation is constructed on $[-5,5]$ with grid size $\Delta \omega = 0.01$. Since $f$ in \eqref{eq:f_example_2} is not strictly convex, Algorithm~\ref{alg:2} was used with initial densities $q_1 = p_{\mathcal{N}}(-0.5,1)$, $q_2 = p_{\mathcal{N}}(0.5,1)$. It reached the required tolerance of $\varepsilon = 10^{-7}$ after $57$ iterations. The resulting least favorable densities are depicted in Fig.~\ref{fig:cost_min_lfds}. Their effect on the cost function can be seen in Fig.~\ref{fig:minimax_cost}, where $r^*(p_0,p_1)$ is plotted for the least favorable and two Gaussian densities. Interestingly, the shape of the cost function $r^*$ is preserved, despite the ``ragged'' shape of the least favorable densities. Again, the proposed algorithm is able to accurately identify abrupt changes as well as smooth variations in the optimal densities.

\section{Conclusion}
\label{sec:conclusion}

The problem of minimizing convex functionals of probability distributions under density band constraints was analyzed. A system of sufficient and necessary first-order optimality conditions was derived as well as a bound on the gap between the exact minimum and the value of the objective function at a candidate solution. The optimality conditions were used to characterize global minimizers of the constrained optimization problem as solutions of a nonlinear fixed-point equation. Two algorithms were proposed that iteratively solve this equation by following a block coordinate descent strategy until the optimality gap falls below a given tolerance. While the first algorithm proved to be efficient in practice, it is not guaranteed to converge for objective functions that are not strictly convex. This problem was overcome by introducing an additional outer proximal iteration. The modified algorithm was then shown to admit guaranteed convergence properties for all band constraints and all convex objective functions, at the cost of a reduced efficiency. Two examples were given to illustrate how the optimality conditions can be used in practice to derive analytical results and to demonstrate the high efficiency and accuracy of the proposed numerical algorithms.


\appendices  


%

\section{Proof of Theorem \ref{th:optimality_condition}}
\label{apx:proof_of_optimality_condition}

Theorem~\ref{th:optimality_condition} is proven by showing that if the densities $(q_1,\ldots,q_N)$ satisfy the conditions in Theorem~\ref{th:optimality_condition}, they also satisfy the Karush--Kuhn--Tucker (KKT) conditions of \eqref{prob:min_explicit}. Since $I_f$ is convex in $(p_1,\ldots,p_N)$ by assumption and the constraints in \eqref{prob:min_explicit} are linear, the KKT conditions are necessary and sufficient for $(q_1,\ldots,q_N)$ to be a global minimizer \cite{Guignard1969}.

The proof makes use of some basic results of infinite-dimensional optimization theory, more precisely, the theory of Lagrange multipliers on Banach spaces. An introduction to the topic is beyond the scope of this paper. A comprehensive treatment can be found, for example, in \cite{Barbu2012}. An elegant standalone proof for the method of Lagrange multipliers and the sufficiency of the KKT conditions is given in \cite{Brezhneva2012} and \cite{Brezhneva2011}, respectively. In brief, the method of Lagrange multipliers can be applied to convex functions on Banach spaces by using Fr\'{e}chet subderivatives instead of subgradients and elements of the dual space instead of scalar- or vector-valued Lagrange multipliers.

Let 
\begin{equation*}
  \mathcal{L}_{\mu}^{\infty} = \left\{ u\colon \Omega \to \mathbb{R} : 
  \sup_{\substack{B \in \mathcal{F} \\ \mu(B) > 0}} \, \sup_{\omega \in B} \; \lvert u(\omega) \rvert < \infty \right\}
\end{equation*}
denote the space of all essentially bounded functions on $\Omega$, which is the dual space of $\mathcal{L}_{\mu}^1$. The Lagrange function $L\colon (\mathcal{L}_{\mu}^1 \times \mathcal{L}_{\mu}^{\infty} \times \mathcal{L}_{\mu}^{\infty} \times \mathbb{R})^N \to \mathbb{R}$ of \eqref{prob:min_explicit} is given by
\begin{equation}
  L(\bm{p},\bm{u},\bm{v},\bm{c}) =  \int_{\Omega} \left( f(\bm{p}) + \sum_{n=1}^N w_n(p_n) \right) \dint\mu + \sum_{n=1}^N c_n,
  \label{eq:lagrange_function}
\end{equation}
where
\begin{equation}
  w_n(p_n) \coloneqq (p_n-\pmax_n)u_n-(p_n-\pmin_n)v_n - p_n c_n
  \label{eq:w}
\end{equation}
and $u_n, v_n \in \mathcal{L}_{\mu}^{\infty}$, $c_n \in \mathbb{R}$ denote the Lagrange multipliers corresponding to the constraints $p_n-\pmax \leq 0$, $p_n-\pmin \geq 0$ and $\int_{\Omega} p_n \dint\mu = 1$, respectively. The dual problem is given by
\begin{gather}
  \max_{\subalign{\{ u_n,v_n &\in \mathcal{L}_{\mu}^{\infty}\}_{n=1}^N \\ \bm{c} &\in \mathbb{R}^N}}
  \left\{ \min_{\{p_n \in \mathcal{L}_{\mu}^1\}_{n=1}^N} \; L(\bm{p},\bm{u},\bm{v},\bm{c}) \right\} 
  \label{eq:dual_problem} \\
  \quad \text{s.t.} \quad u_n, v_n \geq 0, \quad n = 1,\ldots,N\notag
\end{gather}
The partial Fr\'{e}chet-subdifferential of $L$ with respect to $p_n$ can be shown to be
\begin{equation*}
  \partial_{p_n} L (\bm{p},\bm{u},\bm{v},\bm{c}) = \partial_{x_n} f(\bm{p}) + u_n - v_n - c_n.
\end{equation*}
The KKT conditions for the optimal densities require that for all $n = 1,\ldots,N$
\begin{align}
  f_n(\bm{q}) + u_n - v_n - c_n &= 0  & &\text{(stationarity)}
  \label{eq:stationarity} \\
  \pmin_n \leq q_n \leq \pmax_n, \; \int_{\Omega} q_n \dint \mu &= 1 & &\text{(primal feasibility)} 
  \label{eq:primal_feasibility}\\
  u_n, v_n &\geq 0 & &\text{(dual feasibility)}
  \label{eq:dual feasibility} \\
  (q_n-\pmax_n)u_n = (q_n-\pmin_n)v_n &= 0 & &\text{(compl. slackness)}
  \label{eq:compl_slackness}
\end{align}
Let all $q_n$ and $c_n$ be chosen such that they comply with the conditions in Theorem~\ref{th:optimality_condition}. By construction, this implies that $q_n$ satisfies the primal feasibility constraints. Since $f$ is convex and $\mu$ is continuous, it follows from Rademacher's theorem \cite[Chapter 9.J.]{Rockafellar1998} that the partial subderivatives of $f$ are finite $\mu$-almost everywhere, i.e., $f_n \in \mathcal{L}^{\infty}$ for all $n$. Without violating dual feasibility, the functions $v_n$ and $u_n$ can be chosen as
\begin{align}
  -u_n &= ( f_n(\bm{q}) - c_n )^-,  
  \label{eq:lagrange_u} \\
  v_n &= ( f_n(\bm{q}) - c_n )^+, 
  \label{eq:lagrange_v}
\end{align}
so that
\begin{equation}
  v_n-u_n = f_n(\bm{q})-c_n. 
  \label{eq:uv_functions}
\end{equation}
Inserting \eqref{eq:uv_functions} back into the stationarity condition yields
\begin{equation*}
  f_n(\bm{q}) + u_n - v_n - c_n = 0
\end{equation*}
for all $n = 1,\ldots,N$. The last step in the proof is to show that these choices for $q_n$, $u_n$, and $v_n$ also satisfy the complementary slackness constraints, i.e.,
\begin{gather*}
  u_n(\omega) > 0 \; \Rightarrow \; q_n(\omega) = \pmax_n(\omega), \\
  v_n(\omega) > 0 \; \Rightarrow \; q_n(\omega) = \pmin_n(\omega),
\end{gather*}
for all $\omega \in \Omega$. By construction of $u_n$ and $v_n$, $u_n(\omega) > 0$ implies $f_n(\bm{q}(\omega)) < c_n$, which in turn implies $q_n = \pmax_n$. Analogously, $v_n(\omega) > 0$ implies $f_n(\bm{q}(\omega)) > c_n$ and in turn $q_n(\omega) = \pmin_n(\omega)$. $\hfill \square$

\section{Proof of Corollary \ref{cl:optimality_condition}}
\label{apx:proof_of_corollary}

Corollary~\ref{cl:optimality_condition} is a consequence of the fact that $f_n$, being a subderivative of a convex function, is nondecreasing in $x_n$. For the three cases in Theorem~\ref{th:optimality_condition}, it then follows that
\begin{align*}
  f_n(\bm{q}(\omega)) < c_n \; &\Rightarrow \; q_n(\omega) = \pmax_n(\omega) \leq f_n^{-1}(\remidx{q}{n}(\omega),c_n),\\
  f_n(\bm{q}(\omega)) = c_n \; &\Rightarrow \; q_n(\omega) = f_n^{-1}(\remidx{q}{n}(\omega),c_n), \\
  f_n(\bm{q}(\omega)) > c_n \; &\Rightarrow \; q_n(\omega) = \pmin_n(\omega) \geq f_n^{-1}(\remidx{q}{n}(\omega),c_n).
\end{align*}
In words, the three equations state that:
\begin{itemize}
  \item $q_n$ equals its upper bound $p_n''$, if $f_n^{-1}(\bm{q}_{[n]},c_n)$ is larger than $p_n''$;
  \item $q_n$ equals its lower bound $p_n'$, if $f_n^{-1}(\bm{q}_{[n]},c_n)$ is smaller than $p_n'$;
  \item $q_n$ equals $f_n^{-1}(\bm{q}_{[n]},c_n)$ otherwise.
\end{itemize}
The expression for $q_n$ given in Corollary~\ref{cl:optimality_condition} is merely a more compact way of writing this case-by-case definition. $\hfill \square$

\section{Proof of Corollary~\ref{th:optimality_gap}}
\label{apx:proof_of_optimality_gap}

The bound on the optimality gap in Corollary~\ref{th:optimality_gap} can be obtained from the Lagrange dual in \eqref{eq:dual_problem}. By construction, it holds that
\begin{equation}
  \min_{\{p_n \in \mathcal{L}_{\mu}^1\}_{n=1}^N} \; L(\bm{p},\bm{u},\bm{v},\bm{c}) \leq I_f(\bm{q})
  \label{eq:dual_function}
\end{equation}
for all feasible Lagrange multipliers $\bm{u}$, $\bm{v}$, and $\bm{c}$. This minimization is hard to solve in general, but a lower bound on $I_f(\bm{q})$ can be obtained as follows: instead of optimizing over $\bm{p}$, the Lagrange multipliers $\bm{u}$, $\bm{v}$ and $\bm{c}$ can be chosen such that a given $\bm{p}$ satisfies the stationarity conditions in \eqref{eq:stationarity} and, consequently, solves the minimization in \eqref{eq:dual_function}. Let $\bm{u}^*$ and $\bm{v}^*$ denote multipliers that satisfy \eqref{eq:lagrange_u} and \eqref{eq:lagrange_v}, respectively. It is shown in Appendix~\ref{apx:proof_of_optimality_condition} that this choice fulfills the stationarity conditions. It hence holds that
\begin{equation*}
  \min_{\{p_n \in \mathcal{L}_{\mu}^1\}_{n=1}^N} \, L(\bm{p},\bm{u},\bm{v},\bm{c}) = L(\bm{p},\bm{u}^*,\bm{v}^*,\bm{c}).
\end{equation*} 
$L(\bm{p},\bm{u}^*,\bm{v}^*,\bm{c})$ can be shown to evaluate to
\begin{equation}
  L(\bm{p},\bm{u}^*,\bm{v}^*,\bm{c}) = I_f(\bm{p}) - \langle \bm{e}, \bm{1} \rangle,
  \label{eq:objective_lower_bound}
\end{equation}
with $\bm{e}$ defined in Theorem~\ref{th:optimality_gap}. Note that the sum over $c_n$ in \eqref{eq:lagrange_function} cancels with the $p_n c_n$ terms in \eqref{eq:w} since all $p_n$ are assumed to be valid densities. From \eqref{eq:objective_lower_bound}, the bound on the optimality gap follows:
\begin{align*}
  I_f(\bm{p}) - I_f(\bm{q}) &\leq I_f(\bm{p}) - L(\bm{p},\bm{u}^*,\bm{v}^*,\bm{c}) = \langle \bm{e}, \bm{1} \rangle.
\end{align*}
$\hfill \square$

\section{Proof of Convergence of Algorithm~\ref{alg:2}}
\label{apx:proof_of_convergence_algorithm_2}

The convergence of proximal iterations is a well-established result in the convex optimization literature. It follows directly from the contractive property of the proximal operator and can be applied to Algorithm~\ref{alg:2} in a straightforward manner. For a selection of convergence proofs, see, for example, \cite[Chapter 2.3]{Parikh2014}, \cite{Reich2010, Iusem2010, Aoyama2011, Bolte2014}, and the references therein.

In order to prove that Algorithm~\ref{alg:2} converges, it  hence suffices to show that Algorithm 1 indeed solves the inner minimization in \eqref{def:proximal_operator}. This is guaranteed if, first, the function $f$ is strictly convex and, second, if, in every iteration, the equation in line 6 of Algorithm~\ref{alg:1} has a solution. The first condition ensures that the algorithm is able to find a block-coordinate-wise minimum in each iteration; the second condition ensures that the block-coordinate-wise descent indeed converges to a global minimum.

The first condition is fulfilled by construction. Since $\tilde{f}$ is chosen such that it is strictly convex in $(x_1,\ldots,x_n)$, $I_f(p_{\bm{a}_1},\ldots,p_{\bm{a}_N})$ is strictly convex in $(\bm{a}_1,\ldots,\bm{a}_N)$.

The second condition can be shown to be fulfilled for arbitrary convex functions $f$ of the form \eqref{def:f}. By definition of the inverse function in \eqref{eq:f_inverse}, it holds that for every $n = 1, \ldots, N$
\begin{equation*}
  \lim_{c \to -\infty} f_n^{-1}(\remidx{x}{n},c) = 0 \quad \text{and} \quad \lim_{c \to \infty} f_n^{-1}(\remidx{x}{n},c) = \infty
\end{equation*}
for all $\bm{x} \in [0,\infty)^N$. Therefore,
\begin{align*}
  & \lim_{c \to \infty} \langle \min \{ \pmax_n(\bm{\omega}), \max \{ f_n^{-1}(\remidx{A}{n},c), \pmin_n(\bm{\omega}) \} \}, \bm{\mu} \rangle \\
  ={}& \langle \pmax_n(\bm{\omega}), \bm{\mu} \rangle \geq 1
\end{align*}
and
\begin{align*}
  & \lim_{c \to -\infty} \langle \min \{ \pmax_n(\bm{\omega}), \max \{ f_n^{-1}(\remidx{a}{n},c), \pmin_n(\bm{\omega}) \} \}, \bm{\mu} \rangle \\
  ={}& \langle \pmin_n(\bm{\omega}), \bm{\mu} \rangle \leq 1,
\end{align*}
where the last inequalities hold by assumption. Consequently, some $c \in \mathbb{R}$ is guaranteed to exist that solves the equation in line 6 of Algorithm~\ref{alg:1}. $\hfill \square$

\section*{Acknowledgment}

The authors would like to thank the anonymous reviewers for their helpful and constructive comments that greatly improved the quality of the paper.

\ifCLASSOPTIONcaptionsoff
  \newpage
\fi



\bibliographystyle{IEEEtran}
\bibliography{minimizing_convex_functionals}

\begin{thebibliography}{10}
\providecommand{\url}[1]{#1}
\csname url@samestyle\endcsname
\providecommand{\newblock}{\relax}
\providecommand{\bibinfo}[2]{#2}
\providecommand{\BIBentrySTDinterwordspacing}{\spaceskip=0pt\relax}
\providecommand{\BIBentryALTinterwordstretchfactor}{4}
\providecommand{\BIBentryALTinterwordspacing}{\spaceskip=\fontdimen2\font plus
\BIBentryALTinterwordstretchfactor\fontdimen3\font minus
  \fontdimen4\font\relax}
\providecommand{\BIBforeignlanguage}[2]{{%
\expandafter\ifx\csname l@#1\endcsname\relax
\typeout{** WARNING: IEEEtran.bst: No hyphenation pattern has been}%
\typeout{** loaded for the language `#1'. Using the pattern for}%
\typeout{** the default language instead.}%
\else
\language=\csname l@#1\endcsname
\fi
#2}}
\providecommand{\BIBdecl}{\relax}
\BIBdecl

\bibitem{Nguyen2009}
X.~Nguyen, M.~J. Wainwright, and M.~I. Jordan, ``On surrogate loss functions
  and $f$-divergences,'' \emph{The Annals of Statistics}, vol.~37, no.~2, pp.
  876--904, 2009.

\bibitem{Levy2008}
B.~C. Levy, \emph{Principles of Signal Detection and Parameter Estimation},
  1st~ed.\hskip 1em plus 0.5em minus 0.4em\relax New York City, New York, USA:
  Springer, 2008.

\bibitem{Klebanov1978}
L.~B. Klebanov, ``Unbiased estimates and convex loss functions,'' \emph{Journal
  of Soviet Mathematics}, vol.~9, no.~6, pp. 870--880, 1978.

\bibitem{Reid2011}
M.~D. Reid and R.~C. Williamson, ``Information, divergence and risk for binary
  experiments,'' \emph{Journal of Machine Learning Research}, vol.~12, pp.
  731--817, 2011.

\bibitem{KassamPoor1985}
S.~Kassam and H.~Poor, ``Robust techniques for signal processing: A survey,''
  \emph{Proceedings of the IEEE}, vol.~73, no.~3, pp. 433--481, 1985.

\bibitem{Kassam1981}
S.~Kassam, ``Robust hypothesis testing for bounded classes of probability
  densities,'' \emph{IEEE Transactions on Information Theory}, vol.~27, no.~2,
  pp. 242--247, 1981.

\bibitem{Fauss2015}
M.~Fau\ss{} and A.~M. Zoubir, ``Old bands, new tracks---revisiting the band
  model for robust hypothesis testing,'' \emph{IEEE Transactions on Signal
  Processing}, vol.~64, no.~22, pp. 5875--5886, 11 2016.

\bibitem{Fauss2016_thesis}
\BIBentryALTinterwordspacing
M.~Fau{\ss}, ``Design and analysis of optimal and minimax robust sequential
  hypothesis tests,'' Ph.D. dissertation, Technische Universit{\"a}t Darmstadt,
  Darmstadt, Germany, 2016. [Online]. Available:
  \url{http://tuprints.ulb.tu-darmstadt.de/5494/}
\BIBentrySTDinterwordspacing

\bibitem{Rockafellar1968}
R.~T. Rockafellar, ``Integrals which are convex functionals.'' \emph{Pacific
  Journal of Mathematics}, vol.~24, no.~3, pp. 525--539, 1968.

\bibitem{Pearson1900}
K.~Pearson, ``On the criterion that a given system of deviations from the
  probable in the case of a correlated system of variables is such that it can
  be reasonably supposed to have arisen from random sampling,''
  \emph{Philosophical Magazine Series 5}, vol.~50, no. 302, pp. 157--175, 1900.

\bibitem{Mahalanobis1930}
P.~C. Mahalanobis, ``On tests and measures of groups divergence,''
  \emph{Journal of the Asiatic Society of Bengal}, vol.~26, pp. 49--55, 1930.

\bibitem{Shannon1948}
C.~Shannon, ``A mathematical theory of communication,'' \emph{The Bell System
  Technical Journal}, vol.~27, no.~3, pp. 379--423, 1948.

\bibitem{Kullback1951}
S.~Kullback and R.~A. Leibler, ``On information and sufficiency,'' \emph{The
  Annals of Mathematical Statistics}, vol.~22, no.~1, pp. 79--86, 1951.

\bibitem{Liese1987}
F.~Liese and I.~Vajda, \emph{Convex Statistical Distances}.\hskip 1em plus
  0.5em minus 0.4em\relax Leipzig, Germany: Teubner, 1987.

\bibitem{Pardo2005}
L.~Pardo, \emph{Statistical Inference Based on Divergence Measures}.\hskip 1em
  plus 0.5em minus 0.4em\relax Boca Raton, Florida, USA: CRC Press, 2005.

\bibitem{Huber1973}
P.~J. Huber and V.~Strassen, ``Minimax tests and the {N}eyman--{P}earson lemma
  for capacities,'' \emph{The Annals of Statistics}, vol.~1, no.~2, pp.
  251--263, 1973.

\bibitem{Poor1980}
H.~Poor, ``Robust decision design using a distance criterion,'' \emph{IEEE
  Transactions on Information Theory}, vol.~26, no.~5, pp. 575--587, 1980.

\bibitem{Guntuboyina2011}
A.~Guntuboyina, ``Lower bounds for the minimax risk using
  $f$-di\-ver\-gen\-ces, and applications,'' \emph{IEEE Transactions on
  Information Theory}, vol.~57, no.~4, pp. 2386--2399, 2011.

\bibitem{DAmico2014}
A.~A. D'Amico, L.~Sanguinetti, and D.~P. Palomar, ``Convex separable problems
  with linear and box constraints,'' in \emph{Proc. of the IEEE International
  Conference on Acoustics, Speech and Signal Processing (ICASSP)}, 2014, pp.
  5641--5645.

\bibitem{Embrechts2013}
P.~Embrechts and M.~Hofert, ``A note on generalized inverses,''
  \emph{Mathematical Methods of Operations Research}, vol.~77, no.~3, pp.
  423--432, 2013.

\bibitem{Huber1965}
P.~J. Huber, ``A robust version of the probability ratio test,'' \emph{The
  Annals of Mathematical Statistics}, vol.~36, no.~6, pp. 1753--1758, 1965.

\bibitem{Oesterreicher1978}
F.~\"Osterreicher, ``\BIBforeignlanguage{English}{On the construction of least
  favourable pairs of distributions},''
  \emph{\BIBforeignlanguage{English}{Zeitschrift f\"ur
  Wahrscheinlichkeitstheorie und Verwandte Gebiete}}, vol.~43, no.~1, pp.
  49--55, 1978.

\bibitem{AliSilvey1966}
S.~M. Ali and S.~D. Silvey, ``A general class of coefficients of divergence of
  one distribution from another,'' \emph{Journal of the Royal Statistical
  Society. Series B}, vol.~28, no.~1, pp. 131--142, 1966.

\bibitem{GyorfiNemetz1977}
L.~Gy\"orfi and T.~Nemetz, ``$f$-dissimilarity: a general class of separation
  measures of several probability distributions,'' \emph{Colloquia of the
  J\'anos Bolyai Mathematical Society: Topics in Information Theory}, vol.~16,
  pp. 309--321, 1977.

\bibitem{EoM_Approximation_of_Functions}
\BIBentryALTinterwordspacing
M.~Hazewinkel, Ed., \emph{Encyclopaedia of Mathematics}.\hskip 1em plus 0.5em
  minus 0.4em\relax Springer, 2012, vol.~1, ch. Approximation of Functions, pp.
  217--222. [Online]. Available:
  \url{http://www.encyclopediaofmath.org/index.php?title=Approximation_of_functions&oldid=24368}
\BIBentrySTDinterwordspacing

\bibitem{Xu2013}
Y.~Xu and W.~Yin, ``A block coordinate descent method for regularized
  multiconvex optimization with applications to nonnegative tensor
  factorization and completion,'' \emph{SIAM Journal on Imaging Sciences},
  vol.~6, no.~3, pp. 1758--1789, 2013.

\bibitem{Wright2015}
S.~J. Wright, ``Coordinate descent algorithms,'' \emph{Mathematical
  Programming}, vol. 151, no.~1, pp. 3--34, 2015.

\bibitem{Tseng2001}
P.~Tseng, ``Convergence of a block coordinate descent method for
  nondifferentiable minimization,'' \emph{Journal of Optimization Theory and
  Applications}, vol. 109, no.~3, pp. 475--494, 2001.

\bibitem{Parikh2014}
N.~Parikh and S.~Boyd, ``Proximal algorithms,'' \emph{Foundations and Trends in
  Optimization}, vol.~1, no.~3, pp. 127--239, 2014.

\bibitem{Saha2013}
A.~Saha and A.~Tewari, ``On the nonasymptotic convergence of cyclic coordinate
  descent methods,'' \emph{SIAM Journal on Optimization}, vol.~23, no.~1, pp.
  576--601, 2013.

\bibitem{Nutini2015}
J.~Nutini, M.~Schmidt, I.~Laradji, M.~Friedlander, and H.~Koepke, ``Coordinate
  descent converges faster with the {G}auss--{S}outhwell rule than random
  selection,'' in \emph{Proc. of the International Conference on Machine
  Learning (ICML)}, D.~Blei and F.~Bach, Eds., 2015, pp. 1632--1641.

\bibitem{Monahan2001}
J.~Monahan, \emph{Numerical Methods of Statistics}, ser. Cambridge Series in
  Statistical and Probabilistic Mathematics.\hskip 1em plus 0.5em minus
  0.4em\relax Cambridge University Press, 2001, vol.~1.

\bibitem{Waeber2013}
\BIBentryALTinterwordspacing
R.~Waeber, ``Probabilistic bisection search for stochastic root-finding,''
  Ph.D. dissertation, Cornell University, 2013. [Online]. Available:
  \url{https://people.orie.cornell.edu/shane/theses/ThesisRolfWaeber.pdf}
\BIBentrySTDinterwordspacing

\bibitem{Karoui2009}
I.~Karoui, R.~Fablet, J.-M. Boucher, and J.-M. Augustin, ``Seabed segmentation
  using optimized statistics of sonar textures,'' \emph{IEEE Transactions on
  Geoscience and Remote Sensing}, vol.~47, no.~6, pp. 1621--1631, 2009.

\bibitem{Suzuki2004}
H.~Suzuki, R.~Saito, and M.~Tomita, ``The weighted sum of relative entropy: a
  new index for synonymous codon usage bias,'' \emph{Gene}, vol. 335, pp.
  19--23, 2004.

\bibitem{Bhattacharyya1943}
A.~Bhattacharyya, ``On a measure of divergence between two statistical
  populations defined by their probability distributions,'' \emph{Bulletin of
  Calcutta Mathematical Society}, vol.~35, pp. 99--109, 1943.

\bibitem{Domahidi2013}
A.~Domahidi, E.~Chu, and S.~Boyd, ``{ECOS}: {A}n {SOCP} solver for embedded
  systems,'' in \emph{Proc. of the European Control Conference (ECC)}, 2013,
  pp. 3071--3076.

\bibitem{cvx}
{CVX Research, Inc.}, ``{CVX}: Matlab software for disciplined convex
  programming, version 3.0 beta,'' \url{http://cvxr.com/cvx}, 2012.

\bibitem{Guignard1969}
M.~Guignard, ``Generalized {K}uhn--{T}ucker conditions for mathematical
  programming problems in a {B}anach space,'' \emph{SIAM Journal on Control},
  vol.~7, no.~2, pp. 232--241, 1969.

\bibitem{Barbu2012}
V.~Barbu and T.~Precupanu, \emph{Convexity and Optimization in Banach Spaces},
  4th~ed., ser. Springer Monographs in Mathematics.\hskip 1em plus 0.5em minus
  0.4em\relax Houten, Netherlands: Springer, 2012.

\bibitem{Brezhneva2012}
O.~Brezhneva and A.~A. Tret'yakov, ``An elementary proof of the {L}agrange
  multiplier theorem in normed linear spaces,'' \emph{Optimization}, vol.~61,
  no.~12, pp. 1511--1517, 2012.

\bibitem{Brezhneva2011}
------, ``An elementary proof of the {K}arush--{K}uhn--{T}ucker theorem in
  normed linear spaces for problems with a finite number of inequality
  constraints,'' \emph{Optimization}, vol.~60, no.~5, pp. 613--618, 2011.

\bibitem{Rockafellar1998}
R.~T. Rockafellar and R.~J.-B. Wets, \emph{{V}ariational {A}nalysis}.\hskip 1em
  plus 0.5em minus 0.4em\relax New York City, New York, USA: Springer, 1998.

\bibitem{Reich2010}
S.~Reich and S.~Sabach, ``Two strong convergence theorems for a proximal method
  in reflexive {B}anach spaces,'' \emph{Numerical Functional Analysis and
  Optimization}, vol.~31, no.~1, pp. 22--44, 2010.

\bibitem{Iusem2010}
A.~N. Iusem and E.~Resmerita, ``A proximal point method in nonreflexive
  {B}anach spaces,'' \emph{Set-Valued and Variational Analysis}, vol.~18,
  no.~1, pp. 109--120, 2010.

\bibitem{Aoyama2011}
K.~Aoyama, F.~Kohsaka, and W.~Takahashi, ``Proximal point methods for monotone
  operators in {B}anach spaces,'' \emph{Taiwanese Journal of Mathematics},
  vol.~15, no.~1, pp. 259--281, 2011.

\bibitem{Bolte2014}
J.~Bolte, S.~Sabach, and M.~Teboulle, ``Proximal alternating linearized
  minimization for nonconvex and nonsmooth problems,'' \emph{Mathematical
  Programming}, vol. 146, no.~1, pp. 459--494, 2014.

\end{thebibliography}
\end{document}